\magnification=1200 
\vsize=187mm 
\hsize=125mm 

\hoffset=4mm
\voffset=10mm


\abovedisplayskip=4.5pt plus 1pt minus 3pt
\abovedisplayshortskip=0pt plus 1pt
\belowdisplayskip=4.5pt plus 1pt minus 3pt
\belowdisplayshortskip=2.5pt plus 1pt minus 1.5pt
\smallskipamount=2pt plus 1pt minus 1pt
\medskipamount=4pt plus 2pt minus 1pt
\bigskipamount=9pt plus 3pt minus 3pt



\newif\ifpagetitre          \pagetitretrue
\newtoks\hautpagetitre     \hautpagetitre={\hfil}
\newtoks\baspagetitre     \baspagetitre={\hfil\tenrm\folio\hfil}

\newtoks\auteurcourant     \auteurcourant={\hfil}
\newtoks\titrecourant     \titrecourant={\hfil}
\newtoks\chapcourant     \chapcourant={\hfil}

\newtoks\hautpagegauche     \newtoks\hautpagedroite
\hautpagegauche={\vbox{\it\noindent\the\chapcourant\hfill\the\auteurcourant\hfill
{ }\smallskip\smallskip\vskip 2mm\line{}}}
\hautpagedroite={\vbox{\hfill\it\the\titrecourant\hfill{ }
\smallskip\smallskip\vskip 2mm\line{}}}

\newtoks\baspagegauche     \newtoks\baspagedroite
\baspagegauche={\hfil\tenrm\folio\hfil}
\baspagedroite={\hfil\tenrm\folio\hfil}

\headline={\ifpagetitre\the\hautpagetitre
\else\ifodd\pageno\the\hautpagedroite
\else\the\hautpagegauche\fi\fi}

\footline={\ifpagetitre\the\baspagetitre
\global\pagetitrefalse
\else\ifodd\pageno\the\baspagedroite
\else\the\baspagegauche\fi\fi}


\auteurcourant={}
\titrecourant={}


\font\timedouze=cmr12

\font\bfdouze=cmbx12 scaled\magstep1
\font\bfonze=cmbx10 scaled 1100

\def\NN{{\mathord{I\!\! N}}}
\def\RR{{\mathord{I\!\! R}}}
\def\EE{{\mathord{I\!\! E}}}

\def\CC{{\mathord{C\mkern-16mu{\phantom t\vrule}{\phantom o}}}}
\def\LL{{\mathord{I\!\! L}}}
\def\bLL{{\bf\LL}}
\def\TF{\rm{T} \! \Phi} 

\def\bar#1{{\overline{#1}}}
\def\Rp{\RR^+}
\def\pro#1{{(#1_t)}_{t\geq 0}}
\def\norme#1{\left\vert\left\vert #1\right\vert\right\vert}
\def\normca#1{{\left\vert\left\vert #1\right\vert\right\vert}^2}
\def\ab#1{\left\vert #1\right\vert}

\def\titredeux#1#2#3{\centerline{\bfdouze#1}\medskip
     \centerline{\bfdouze#2}\bigskip\bigskip\bigskip\centerline{\timedouze#3}}

\def\spa#1#2{\bigskip\medskip\noindent{\bfdouze #1\ #2}\par\nobreak}
\def\sspa#1#2{\bigskip\noindent{\bfonze #1\ #2}\par\nobreak\smallskip}

\def\th#1{\bigskip\smallskip\noindent{\bf Theorem #1}$\,$--$\,$}

\def\thl#1#2{\bigskip\smallskip\noindent{\bf Theorem #1} #2$\,$--$\,$}
\def\prp#1{\bigskip\smallskip\noindent{\bf Proposition #1}$\,$--$\,$}

\def\le#1{\bigskip\smallskip\noindent{\bf Lemma #1}$\,$--$\,$}

\def\co#1{\bigskip\smallskip\noindent{\bf Corollary #1}$\,$--$\,$}

\def\prf{\bigskip\noindent{\bf Proof}\par\nobreak\smallskip}

\def\ecarte{\vphantom{\buildrel\bigtriangleup\over =}}
\def\findem{\hfill\hbox{\vrule height 2.5mm depth 0 mm width 2.5 mm}}
\def\indic{{\mathop{\rm 1\mkern-4mu l}}}

\def\di{\displaylines}
\def\eq{\eqalignno}
\def\hf{\hfill}
\def\wh{\widehat}
\def\wt{\widetilde}
\def\ps#1#2{{<}\, #1\, ,\, #2\,{>}}
\def\hp{\hphantom}
\def\ld{\ldots}
\def\cd{\cdot}

\def\qed{\findem}

\def\qq{\qquad}

\def\cd{\cdot}

\def\frac#1#2{{{#1}\over{#2}}}
\def\seq#1{{(#1_n)}_{n\in\NN}}

\def\rB{{\cal B}}\def\rD{{\cal D}}
\def\rF{{\cal F}}\def\rH{{\cal H}}
\def\rL{{\cal L}}
\def\rP{{\cal P}}
\def\rS{{\cal S}}

\def\ut{\tilde \phi}
\def\vt{\tilde \psi}

\def\a{\alpha}

\def\d{\delta}
\def\e{\varepsilon}
\def\f{\phi}
\def\s{\sigma}

\def\r{\rho}

\def\x{\chi}

\def\o{\omega}

\def\G{\Gamma}

\def\L{\Lambda}

\def\O{\Omega}

\def\Y{\Psi}
\def\N{\nabla}

\def\F{\Phi}

\def\NNE{{\NN^\ast}}

\auteurcourant={St\'ephane ATTAL and Yan PAUTRAT}
\titrecourant={From repeated to continuous quantum interactions}

\titredeux{From repeated to continuous}{quantum
interactions}{St\'ephane ATTAL and Yan PAUTRAT} 
\bigskip
\bigskip
\sspa{}{Abstract}{\sevenrm We consider the general physical situation of a
 quantum system $\rH_0$ interacting with a chain of exterior
systems $\otimes_\NNE \rH$, one after the other, during a small interval
of time $h$ and following some Hamiltonian $H$ on
$\rH_0\otimes\rH$. We discuss the passage to the limit to continuous
interactions ($h\rightarrow 0$) in a setup which allows to
compute the limit of this Hamiltonian evolution in a single state space:
a continuous field of exterior systems $\otimes_{\Rp}\rH$. Surprisingly,
the passage
to the limit
shows the necessity for 3 
different time scales in $H$.  The
limit evolution equation is shown to spontaneously produce quantum
noises terms: we obtain a quantum Langevin equation as  limit
of the Hamiltonian evolution. For the very
first time, these quantum Langevin equations are obtained as the
effective limit from repeated to continuous interactions and not only
as a model. These results justify the usual quantum Langevin equations
considered in continual quantum measurement or in quantum optics.  We
show that the three time scales  correspond to the 
normal regime, the weak coupling limit and the low density limit. Our
approach allows to consider these two physical limits
altogether for the first time. Their combination produces an
effective Hamiltonian on the small system, which had never been
described before. We apply these results to give an Hamiltonian
description of the von Neumann measurement. We
also consider the approximation of  
continuous time quantum master equations by discrete time ones. In
particular we show how any Lindblad
generator is obtained as the limit of completely positive maps.}

\def\norme#1{\left\| #1 \right\|} 
\def\normeca#1{{\left\| #1 \right\|}^2} 
\def\normei#1{\left\| #1 \right\| _{\infty}}

\def\somu#1{\sum_{\scriptstyle{#1}}}  
\def\conj#1{\overline{#1}} 
  
\def\intk#1{\int _{t_#1} ^{t_{#1+1}} \!\!}
\vfill\eject
\spa{}{Contents}
\bigskip
\noindent I. {\bf Introduction}

\smallskip
\noindent II. {\bf Discrete dynamics on the atom chain}

II.1 Repeated quantum interactions

II.2 Structure of the atom chain

II.3 Unitary dilations of completely positive maps

\smallskip
\noindent III. {\bf From the atom chain to the atom field}

III.1 Structure of the atom field

III.2 Quantum noises

III.3 Embedding and approximation by the atom chain

III.4 Quantum Langevin equations

\smallskip
\noindent IV. {\bf Convergence theorems}

IV.1 Convergence to quantum Langevin equations

IV.2 Typical Hamiltonian: weak coupling and low density

IV.3 Hamiltonian description of von Neumann measurements

IV.4 One example

IV.5 From completely positive maps to Lindbladians

\bigskip
\bigskip
\spa{I.}{Introduction}

Quantum Langevin equations as a model for quantum open systems have
been considered for at least 40 years (for example [FKM], [FLO],
[AFL]). They have been given many 
different meanings in terms of several definitions of quantum noises
or quantum Brownian motions (for example [G-Z], [H-P], [GSI]). One of the most developed
and useful mathematical languages developed for that purpose is the
quantum stochastic calculus of Hudson and Parthasarathy and their
quantum stochastic differential equations ([H-P]). The quantum
Langevin equations they allow to consider have been used very often to
modelize typical situations of quantum open systems: continual quantum
measurement ([Ba1], [B-B]), quantum optics ([F-R], [FRS]
[Ba2]), electronic transport [BRSW], thermalization ([M-R], [L-M]),
etc.

The justification for such quantum Langevin equation is often given in
terms of some particular approximations  of the true  
Hamiltonian interaction dynamic: rotating wave approximation,
Markov approximation, large band approximation  (cf [G-Z] chapter 11).

They are also often justified as natural dilations of quantum
master equations on the small system. That is, for any (good) semigroup 
of completely positive maps on the small system (with Lindblad
generator $\rL$), one can dilate the small system with an appropriate 
Fock space, and obtain an explicit quantum stochastic differential equation
on the whole space. The unique solution of this equation
is a unitary evolution (in interaction picture) such that the trace on the small system of the
induced  evolution yields the original semigroup. This
corresponds, at the quantum level, to the well-known way of realizing a
concrete Markov process from a given semigroup (or 
generator) by adding a noise space to the (classical) system space and
solving an adequate stochastic differential equation. 

Some quantum stochastic differential equations  have also been
obtained in the so-called {\it stochastic limit} 
from explicit Hamiltonian dynamics ([A-L], [AGL], [ALV]). This
shows some similarities with the results described here, but
the limits considered in those articles are in the sense of the
convergence of processes living in a different 
space than the one of the Hamiltonian dynamic.
\bigskip
In this article we consider the effective Hamiltonian dynamic
describing the repeated interactions, during short time intervals of length
$h$, of a small system $\rH_0$ with a chain of exterior systems
$\otimes_\NNE \rH$. We embed all these chains as particular subspaces,
attached to the parameter $h$,  
of a continuous field 
$$\bigotimes_{\Rp} \rH$$
in such a way that the subspaces associated to the chain increase and fill the
field when $h$ tends to 0. This framework may seem to specialize to
the case of a zero-temperature exterior system; actually, as was noted
by the first author and Maassen, it also applies to the case of
positive temperature, using the cyclic (GNS) representation of the
given state. 

By developing an appropriate language of the chain $\otimes_\NNE\rH$ and of
the field $\otimes_{\Rp}\rH$ and by describing the discrete time
Hamiltonan evolution generated by the repeated interactions, we are
able to pass to the limit when $h\rightarrow 0$ and prove that the limit
evolution operator is the solution of a quantum stochastic differential
equation. This limit is obtained in the strong topology of operators
in a single space:  the continuous field $\bigotimes_{\Rp}\rH$, and
implies the weak convergence of the Heisenberg evolutions of any
observable.

\bigskip
Of course, such a limit cannot be obtained without assumptions on the
elementary interaction Hamiltonian $H$. This is similar to the central
limit theorem: a random walk gives a trivial limit when its time step
$h$ goes to zero and it is only when suitably renormalized (by a
factor $\sqrt h$) that it yields a Gaussian. Other normalizations give
either trivial limits or no limit at all. 

In our Hamiltonian context the situation is going to be the same. For
a non-trivial limit of these repeated interactions to exist, we will
need the Hamiltonian $H$ to satisfy some renormalization
properties. The surprise here is that the necessary renormalization
factor is not global, it is different following some parts of the
Hamiltonian operator. We identify 3 different time scales in $H$: one
of order 1, one of order $\sqrt h$, one of order $h$. 

We describe a class of Hamiltonian which seems to be typical for the above
conditions to be satisfied. These typical Hamiltonians are clearly a
combination of free evolution, weak coupling limit typical
hamiltonians and low density limit typical Hamiltonians. This physically
explains the three different time scales. But the originality of our
approach allows to consider both limits together; to our knowledge
this constitutes 
a novelty in the literature. As a consequence, the
combination of the two limits shows an effective Hamiltonian for
the small system which is very surprising:  it contains a new term 
$$
V^\ast D^{-2}(\sin D-D)V
$$
which comes from the presence of both the weak coupling and the low
density limit in the Hamiltonian. It seems that such a term had never
been described before. Again notice a possible extension of our
results: only the case of time-independent coupling is discussed here
but results for time-dependent ones can easily be deduced.

\bigskip
This article is structured as follows. 

In section II we present the
exact mathematical model of repeated quantum interactions and end up with
the associated evolution equation (subsection II.1). We then introduce
a mathematical setup for the study of the space $\bigotimes_\NNE\rH$ which
will help much for passing to the continuous field. In particular
this includes a particular choice of an orthonormal basis of the phase
space and a particular choice of a basis for the operators on that
phase space (subsection II.2). Finally we show how the typical
evolution equations obtained in II.1 are the general model for the
unitary dilation of any given discrete semigroup of completely
positive maps (subsection II.3).

Section III is devoted to presenting the whole formalism of the
continuous atom field. In subsection III.1 we present the space which
is candidate for representing the continuous field limit of the atom
chain. It is actually a particular Fock space on which we develop an
unusual structure which clearly shows the required properties. In
subsection III.2 we present the natural quantum noises on the
continuous field and the associated quantum stochastic integrals, the
quantum Ito formula and the quantum stochastic differential equations.
In subsection III.3 we concretely realize the atom chain of section II
as a strict subspace of the atom field. Not only do we realize it as a
subspace, but also realize the action of its basic operators inside
the atom field. All these atom chain subspaces are related to a
partition of $\Rp$. When the diameter of the partition goes to 0, we
show that the corresponding subspace completely fills the
continuous field and the basic operators of the chain converge to the
quantum noises of the field (with convenient normalizations). Finally,
considering the projection of the continuous atom field onto an atom
chain subspace, we state a formula for the projection of a general quantum
stochastic integral.

In section IV all the pieces of the puzzle fit together. By computing
the projection on the atom chain of a quantum stochastic differential
equation we show that the typical  evolution equation of repeated
interactions converges in the field space to the solution of a
quantum Langevin equation, assuming the fact that the associated
Hamiltonian satisfies some particular renormalization property
corresponding to three different time scales. It is to that result
and to some of its extensions
that subsection IV.1 is devoted. In subsection IV.2 we describe a
family of Hamiltonians which seems to be typical of the conditions
obtained above. We show that this family of Hamiltonians describes
altogether free evolution, weak coupling limit and low density limit 
terms. Computing the associated quantum Langevin equation at the limit,
we obtain an effective Hamiltonian on $\rH_0$ which contains a new
term. This new term appears only when weak coupling and low density
limits are in presence together in the Hamiltonian. In subsection
IV.3, we apply these results to describe the von Neumann measurement
apparatus in the 
Hamiltonian framework of repeated quantum interactions. In subsection IV.4 we
explicitly
compute a simple example. In subsection IV.5 we show that our
approximation theorem 
puts into evidence a natural way that completely positive maps have to
converge to Lindblad generators.

\spa{II.}{Discrete dynamics on the atom chain}

\sspa{II.1}{Repeated quantum interactions}

We here give a precise description of our physical model: repeated
quantum interactions. 
\bigskip
We consider a small quantum system $\rH_0$ and another quantum system $\rH$ which
represents a piece of environment, a measuring apparatus or incoming
photons$\ld$ We consider the space $\rH_0\otimes \rH$ in order to
couple the two systems, an Hamiltonian $H$ on $\rH_0\otimes\rH$ which
describes the interaction and the associated unitary evolution during
the interval $[0,h]$ of time:
$$
\LL=e^{-ihH}.
$$
This single interaction is therefore described in the Schr\"odinger
picture by
$$
\r\mapsto \LL\,\r\,\LL^\ast
$$
and  in the Heisenberg picture by
$$ X \mapsto \LL^\ast X \LL.$$

\bigskip
Now, after this first interaction, we repeat it but this time coupling
the same $\rH_0$ with a new copy of $\rH$. This means that that new
copy was kept isolated until then; similarly the previously considered
copy of $\rH$ will remain isolated for the rest of the experience. One 
can think of many physical examples where this situations arises: in
repeated quantum measurement where a family of identical measurement
devices are presented one after the other before the system (or a
single device is refreshed after every use), in quantum optics where
a sequence of independent atoms arrives one after the other to
interact with a field in some cavity for a short time. More generally
it can be seen as a good model if it is assumed that perturbations in
$\rH$ due to the interaction are dissipated after every time $h$. 

The sequence of interactions can be described in the following way: the state space for the whole system is
$$\rH_0\otimes\bigotimes_\NNE\rH
$$
Index for a few lines only the copies of $\rH$ as $\rH _1$, $\rH _2$,
$\ld$ Define then a unitary operator $\LL _n$ as the canonical
ampliation to $\rH_0\otimes \rH 
_1 \otimes \rH _2 \otimes \ldots$ of the operator which acts as $\LL$
on $\rH _0 \otimes \rH_n$; that is, $\LL _n$ acts as the identity on
copies of $\rH$ other than $\rH _n$. 

The effect of the n-th interaction in the Schr\"odinger picture writes then
$$ \r \mapsto \LL _n \, \r\, \LL _n ^*,$$
for every density matrix $\r$, so that the effect of the $n$ first interactions is
$$ \r \mapsto u_n\, \r\, u_n^*$$
where $(u_n)_{n\in\NN}$ is a sequence in $\rB(\rH _0 \otimes \bigotimes _{\NNE}\rH)$ which satisfies the equations
$$  \cases{ u_{n+1} = \LL_{n+1} \, u_n \cr \hp{{}_{1}}u_0 \hp{{}_{+}}
= \ I.} \eqno{(1)}
$$
It is evolution equations such as (1) that we are going to study in this article.

\sspa{II.2}{Structure of the atom chain}

We here describe some useful mathematical structure on the space
$\otimes_\NNE\rH$ which will constitute the main ingredient of our
approach.
\bigskip
Let us fix a particular Hilbertian basis $(X^i)_{i\in\L \cup \{0\}}$
for the Hilbert space $\rH$, where we assume (for notational purposes)
that $0 \not\in \L$. This particular choice of notations is motivated
by physical  interpretations: indeed, we see the
$X^i$, $i \in \L$, as representing for example the different possible
excited states of an atom. The vector $X^0$ represents the ``ground
state" or ``vacuum state" of the atom and will usually be denoted
$\Omega$. 

Let $\TF$ be the tensor product $\bigotimes _{\NNE} \rH$ with respect
to the stabilizing sequence $\Omega$. In other words, this means
simply that an orthonormal basis of $\TF$ is given by the family
 
$$ \{ X_A; \ A \in \rP_{\NNE,\L}\}$$
where
\def\seqe#1{{{(#1_n)}_{n\in\NNE}}}

-- the set $\rP _{\NN, \L}$ is  the set of finite subsets 
$$ \{ (n_1, i _1), \ldots, (n_k, i_k)\}$$
of $\NNE \times \L$ such that the $n_i$'s are mutually different.
Another way to describe the set $\rP_{\NNE,\L}$ is to identify
   it to the set of sequences $\seqe A$ with values in $\L \cup \{0\}$
   which take a value different from 0 only  finitely often.

-- $X_A$ denotes the vector 
$$
\O\otimes\ld\otimes\O\otimes X^{i_1}\otimes
\O\otimes\ld\otimes\O\otimes X^{i_2}\otimes\ld
$$
where $X^{i_1}$ appears in $n_1$-th copy of $\rH$...

\smallskip
The physical signification of this basis is easy to understand: we
have a chain of atoms, indexed by $\NNE$. The space $\TF$ is the state
space of this chain, the vector $X_A$ with $A=\{ (n_1, i_1), \ldots,
(n_k, i _k)\}$ representing the state in which exactly $k$ atoms are
excited: atom $n_1$ in the state $i _1$, etc, all other atoms being in
the ground state.
 
\bigskip 
This particular choice of a basis gives $\TF$ a particular
structure. If we denote by $\TF_{n]}$ the space generated by the $X_A$
such that $A\subset\{1,\ld,n\}\times\L$ and by $\TF_{[m}$ the one generated by
the $X_A$
such that $A\subset\{m,m+1,\ld\}\times\L$, we get an obvious natural
isomorphism between $\TF$ and $\TF_{n-1]}\otimes \TF_{[n}$ given by
$$
[f\otimes g](A)=f\left(A\cap\{1,\ld,n-1\}\times\L\right)\,
g\left(A\cap\{n,\ld\}\times\L\right).
$$

Put $\{a^i_j;i,j \in \L\cup\{0\} \}$ to be the natural basis of $\rB(\rH)$,
that is,
$$
a^i_j(X^k)=\d_{ik}X^j.
$$
We denote by $a^i_j(n)$ the natural ampliation of the operator $a^i_j$
to $\TF$ which acts on the copy number $n$ as $a^i_j$ and the identity
elsewhere. That is, in terms of the basis $X_A$, 
$$
a^i_j(n)X_A=\indic_{(n,i)\in A}X_{A\setminus (n,i)\cup (n,j)}
$$
if neither $i$ nor $j$ is zero, and
$$
\eqalign{
a^i_0 (n) X_A &= \indic_{(n,i) \in A} X_{A \setminus (n,i)}, \cr
a^0_j(n)X_A &=\indic_{(n,0) \in A} X_{A\cup (n,j)},\cr
a^0_0 (n) X_A &= \indic_{(n,0) \in A} X_{A},\cr}
$$
where $(n,0)\in A$ actually means ``for any $i$ in $\L$,
$(n,i)\not\in\L$''.
\bigskip

\sspa{II.3}{Unitary dilation of completely positive maps}

The evolution equations
$$
u_n=\LL_n\ld \LL_1
$$
obtained in the physical setup of repeated quantum interactions are
actually of mathematical interest on their own for they provide a
canonical way of dilating discrete semigroups of completely positive
maps into unitary automorphisms.
\bigskip
The mathematical setup is the same. Let $\LL$ be any operator on
$\rH_0\otimes\rH$. Let $\TF=\otimes_{\NN^\ast}\rH$ and $(\LL_n)_{n\in\NNE}$ be
defined as in the above section. 
We then consider the associated evolution equations
$$
u_n=\LL_n\ld \LL_1\eqno{(1)}
$$
with $u_0=I$. 
\bigskip
The following result is obvious.

\prp{1.}{\it The solution $\seq u$ of (1) is made of unitary
(resp. isometric, contractive) operators if and only if $\LL$ is unitary
(resp. isometric, contractive).}

\qed
\bigskip
Note that if $\LL$ is unitary, then the mappings 
$$
j_n(H)=u_n^\ast H u_n
$$
are automorphisms of $\rB(\rH_0\otimes\rH)$. 
\bigskip
Let $\EE_0$ be the partial trace on $\rH_0$ defined by
$$
\ps{\f}{\EE_0(H)\,\psi}=\ps{\f\otimes\O}{H\,\psi\otimes\O}
$$
for all $\f,\psi\in\rH_0$ and every operator $H$ on $\rH_0\otimes \TF$. 

Unitary dilations of completely positive semigroups are obtained in
the following theorem. Recall that, by Kraus' theorem, any completely
positive operator $\ell$ on $\rB(\rH _0)$ is of the form
$$ \ell(X) = \sum _{i\in \NN} A_i^\ast X A_i$$
where the summation ranges over $(\L \cup \{0\})^2$, the $A_i$ are bounded operators and the sum is strongly
convergent. Conversely, any such operator is completely positive.
\smallskip
\noindent{\bf Remark}: Of course the Kraus form of an operator is {\it
a priori} indifferent to the specificity of the value $i=0$. The
special role played by one of the indices will appear later on. 

\th{2.}{\it Let $\LL$ be any unitary operator on
$\rH_0\otimes\rH$. Consider the coefficients
$(\LL^i_j)_{i,j\in\L\cup\{0\}}$, which are operators on $\rH_0$, of the
matrix representation of $\LL$ in the basis  $\O,X^i$, $i\in\L$ of $\rH$. 

Then, for any $X\in\rB(\rH_0)$ we have
$$
\EE_0[j_n(X\otimes I)]=\ell^n(X)
$$
where  $\ell$ is the completely positive map on $\rB(\rH_0)$ given by
$$
\ell(X)=\sum_{i\in\L\cup\{0\}} (\LL^0_i)^\ast X \LL^0_i.
$$
Conversely, consider any completely positive map
$$
\ell(X)=\sum_{i\in\L\cup\{0\}} A_i^\ast XA_i
$$  
 on
$\rB(\rH_0)$ such  that $\ell(I)=I$. Then  there
exists a unitary operator $\LL$ on $\rH_0\otimes\rH$ such
that the associated unitary family of automorphisms
$$
j_n(H)=u_n^\ast Hu_n
$$
satisfies
$$
\EE_0[j_n(X\otimes I)]=\ell^n(X),
$$
for all $n\in\NN$. }

\prf

Consider $\LL=(\LL^i_j))_{i,j\in\L\cup\{0\}}$ such as in the above
statements. Consider the unitary family 
$$
u_n=\LL_n\ld \LL_1.
$$
Note that 
$$
u_{n+1}=\LL_{n+1}u_n.
$$
Put $j_n(H)=u_n^\ast H u_n$ for every  operator $H$ on
$\rH_0\otimes\rH$. Then, for any operator $X$ on $\rH_0$ we have
$$
j_{n+1}(X\otimes I)= u_n^\ast \LL_{n+1}^\ast(X\otimes
I)\LL_{n+1}u_n.
$$
When considered as a  matrix of operators on $\rH_0$, in the
basis $\O,X^i$, $i\in\L$ of $\rH$, the matrix associated to
$X\otimes I$ is of diagonal form.  We  get
$$
\di{
\LL_{n+1}^\ast(X\otimes
I)\LL_{n+1}=\hf\cr
\hf=\left(\matrix{(\LL^0_0)^\ast&(\LL^0_1)^\ast&\ld\cr (\LL^1_0)^\ast\ecarte&(\LL^1_1)^\ast&\ld\cr
\vdots&\vdots&\ddots\cr}\right)
\left(\matrix{X&0&\ld\cr 0&\ecarte X&\ld\cr
\vdots&\vdots&\ddots\cr
}\right)\left(\matrix{\LL^0_0&\LL^1_0&\ld\cr
\LL^0_1&\ecarte\LL^1_1&\ld\cr 
\vdots&\vdots&\ddots}\right)\cr
}
$$
\def\bLL{{\bf\LL}}
which is  the matrix $\bLL_{n+1}(X)={(B^i_j(X))}_{i,j\in\L\cup\{0\}}$ with
$$
B^i_j(X)=\sum_{k\in\L\cup\{0\}} (\LL^j_k)^\ast X\LL^i_k.
$$
Note that the operator $\bLL_{n+1}(X)$ acts non trivialy only on the
tensor product of $\rH_0$ with the 
$(n+1)$-th copy of $\rH$. When represented as an operator on
$$
\rH_0\otimes \TF_{n+1]}=\left(\rH_0\otimes
\TF_{n]}\right)\otimes\rH
$$
as a  matrix with coefficients in $\rB(\rH_0\otimes
\TF_{n]})$ it writes exactly in the same way as above, just replacing  $B^i_j(X)$
(which belongs to $\rB(\rH_0)$) by $$B^i_j(X)\otimes I_{\vert\TF_{n]}}.$$

Also note that, as can be proved by an easy induction,
the operator $u_n$ acts on $\rH_0\otimes \TF_{n]}$ only. As an
operator on $\rH_0\otimes \TF_{n+1]}$ it is represented by a diagonal
matrix.   Thus 
$j_{n+1}(X)= u_n^\ast\bLL_{n+1}(X)u_n$ 
can be written  on $\rH_0\otimes \TF_{(n+1)]}=\rH_0\otimes
\TF_{n]}\otimes\rH$ as a matrix of operators on
$\rH_0\otimes \TF_{n]}$ by
$$
\left(j_{n+1}(X\otimes I)\right)^i_j=j_n(B^i_j(X)\otimes I).
$$
Note that $B^0_0(X)=\sum_{i\in\L\cup\{0\}}(\LL^0_i)^\ast X \LL^0_i$ which is the
mapping $\ell(X)$ of the statement.

Put $T_n(X)=\EE_0[j_n(X\otimes I)]$. We have, for all $\f,\Y\in\rH_0$
$$
\eq{
\ps{\f}{T_{n+1}(X)\Y}&=\ps{\f\otimes\O}{j_{n+1}(X\otimes I)\,\Y\otimes\O}\cr
&=\ps{\f\otimes\O}{\left(j_{n}(B^i_j(X)\otimes
I)\right)_{i,j}\Y\otimes\O}\cr
&=\ps{\f\otimes\O_{\,\TF_{n]}}\otimes\O_{\,\rH}}{\left(j_{n}(B^i_j(X)\otimes
I)\right)_{i,j}\Y\otimes\O_{\,\TF_{n]}}\otimes\O_{\,\rH}}\cr
&=\ps{\f\otimes\O_{\,\TF_{n]}}}{j_{n}(B^0_0(X)\otimes I)\Y\otimes\O_{\,\TF_{n]}}}\cr
&=\ps{\f}{T_n(\ell(X))\Y}.\cr
}
$$
This proves that $T_{n+1}(X)=T_n(\ell(X))$ and the first part of the
theorem is proved.
\smallskip
Conversely, consider a decomposition of a completely positive map $\ell$ of the form
$$
\rL(X)=\sum_{i\in\L\cup\{0\}} A_i^\ast XA_i
$$
for a familly $(A_i)_{i\in\L\cup\{0\}}$ of bounded operators on $\rH_0$ such that
$$\sum_{i\in\L\cup\{0\}} A_i^\ast A_i=I.$$

We claim that there exists a unitary operator $\LL$ on $\rH_0\otimes \rH$
of the form
$$
\LL=\left(\matrix{A_0&\ld&\ld\cr A_1&\ld&\ld\cr\vdots&\vdots&\ddots\cr
}\right).
$$
Indeed, the condition $\sum_{i\in\L\cup\{0\}}A_i^\ast A_i=I$ guarantees that the first 
columns of $\LL$  are made of orthonormal
vectors of $\rH_0\otimes \rH$. We can thus complete the matrix by
completing it into an orthonormal basis of $\rH_0\otimes \rH$. This
makes out a unitary matrix $\LL$ the coefficients of which  we denote by
${(A^i_j)}_{i,j\in\L\cup\{0\}}$. Note that $A^0_i=A_{i+1}$. We now conclude
easily by the first part of the theorem.

\qed

\spa{III}{From the atom chain to the atom field}

\sspa{III.1}{Structure of the atom field}

We now describe the structure of the continuous version of
$\TF$. The  structure we are going to present here
is rather original and not  much expanded in the literature. It
is very different from the usual presentation of quantum stochastic
calculus ([H-P]), but it actually constitutes a very natural
language for our purpose: approximation of the atom field by atom
chains. This approach is taken from [At1].

We first start with a heuristical discussion.

By a continuous version of the atom chain $\TF$ we mean  a Hilbert
space with a structure which makes it 
the space  
$$
\F=\bigotimes_{\RR^+}\rH.
$$
We have to give a meaning to the above notation. This could be
achieved by invoquing  the framework of continous tensor products of Hilbert
spaces (see [Gui]), but we prefer to give a self-contained
presentation which fits better with our approximation procedure.

 Let us make out an
idea of what it should look like by mimicking, in a 
continuous time version, what we have described in $\TF$. 

The countable orthonormal basis $X_A, A\in\rP_{\NNE, \L}$ is replaced by
a continuous orthonormal  basis $d\x_\s,\, \s\in\rP_{\RR, \L}$, where $\rP_{\RR, \L}$ is the
set of finite subsets of $\RR^+\times \L$. With the same idea as for $\TF$,
this means that each copy of $\rH$ is equipped with an orthonormal
basis $\O,d\x^i_t$, $i\in\L$ (where $t$ is the parameter attached to the copy we
are looking at). The orthonormal basis above is just the one 
obtained by specifying a finite number of sites $t_1,\ld,t_n$ which are
going to be excited, the other ones being supposed to be in the
fundamental state $\O$, and by specifying their level of excitation.

The representation of an element $f$ of $\TF$:
$$
\eqalign{
f&=\sum_{A\in\rP_{\NNE, \L}} \! f(A)\, X_A\cr
\normca f&=\sum_{A\in\rP_{\NNE,\L}} \! \ab{f(A)}^2\cr
}
$$
is replaced by an integral version of it in $\F$:
$$
\eq{
f&=\int_{\rP_{\RR,\L}} \! f(\s)\, d\x_\s\cr
\normca f&=\int_{\rP_{\RR,\L}} \! \ab f^2\, d\s.\cr
}
$$
This last integral has to be explained: the measure $d\s$ is  a
``Lebesgue measure'' on $\rP_{\RR,\L}$, as  will be explained later.
From now on, the notation $\rP$ will denote, depending on the context, spaces of the type $\rP_{\NNE,\L}$ or $\rP_{\RR,\L}$.

A good basis of operators acting on $\F$ can be obtained by mimicking
the operators $a^i_j(n)$ of $\TF$. We will here have a set of
infinitesimal operators $da^i_j(t)$, $i,j \in\L\cup\{0\}$,  acting on the ``t-th" copy of
$\rH$ by:
$$
\eq{
da^0_0(t)\, d\x_\s&=d\x_{\s}\, dt\, \indic_{t\not\in \s}\cr
da^0_i(t)\, d\x_\s&=d\x_{\s\cup \{(t,i)\}}\, \indic_{t\not\in \s}\cr
da^i_0(t)\, d\x_\s&=d\x_{\s\setminus \{(t,i)\}}\, dt\, \indic_{(t,i)\in \s}\cr
da^i_j(t)\, d\x_\s&=d\x_{\s\setminus\{(t,i)\}\cup\{(t,j)\}}\,
\indic_{(t,i)\in \s}\cr 
}
$$
for all $i,j \in \L$.
\bigskip
We shall now describe a rigourous setup for the above heuristic discussion.

\def\pcc{\rP}
\def\rb{\RR}
\def\cb{\CC}
\def\pcn{\rP_n}
\def\fc{\rF}
\def\vi{\emptyset}

We recall the structure of the bosonic Fock space $\F$ and its basic
structure (cf [At1] for more details and [At3] for a complete study of
the theory and its connections with classical stochastic processes).

Let $\rH$ be, as before, a Hilbert space with an orthonormal basis $X^i$, $i \in \L \cup \{ 0\}$ and let $\rH '$ be the closed subspace generated by vectors $X^i$, $i\in\L$ (or simply said, the orthogonal of $X^0$). 

Let $\F=\G_s(L^2(\Rp, \rH '))$ be the symmetric (bosonic) Fock space over the space $L^2(\Rp, \rH ')$.
We shall here give a very efficient presentation of that space, the
so-called {\it Guichardet interpretation} of the Fock space. 

Let $\pcc$ ($=\pcc_{\rb, \L}$) be the set of finite subsets
$\{(s_1,i_1),\ld,(s_n,i_n)\}$ of
$\rb^+\times \L$ such that the $s_i$ are two by two different. Then $\pcc = \cup_n \pcn$ where $\pcn$ is the subset of $\pcc$ made of $n$-elements subsets of
$\ \rb^+\times \L$. By ordering the $\Rp$-part of the elements of
$\s\in\pcn$, the set $\pcn$ can be identified to the increasing simplex
$\Sigma _n = \{0<t_1 < \cdots <t_n\}\times \L$ of $\rb^n\times \L$. Thus $\pcn$
inherits a measured space structure from the Lebesgue measure on
$\ \rb^n$ times the counting measure on $\L$. This also gives a measure
structure on $\pcc$ if we specify 
that on $\pcc_0 = \{\vi\}$ we put the measure $\delta
_\vi$. Elements of $\pcc$ are often denoted by $\sigma $, the measure on
$\pcc$ is denoted $d\sigma $. The $\sigma $-field obtained this way on
$\pcc$ is denoted $\fc$.

We identify any element $\s\in\rP$ with a family
$\{\s_i, \,  i\in \L \}$ of (two by two disjoint)  subsets of $\Rp$ where 
$$
\s_i=\{s\in\Rp; (s,i)\in\s\}.
$$

\def\Unn{\indic}

The {\it Fock space\/} $\Phi$ is the space $L^2(\pcc,\fc,d\sigma
)$. An element $f$ of $\Phi$ is thus a measurable function $f:\pcc \to
\cb$ such that
$$
\normca{f} = \int_{\pcc} |f(\sigma )|^2\ d\sigma  < \infty.
$$
One can define, in the same way, $\pcc_{[a,b]}$ and $\Phi_{[a,b]}$ by
replacing $\rb^+$ with $[a,b]\subset \rb^+$. As in discrete time, there is a natural
isomorphism between $\Phi_{[0,t]} \otimes  \Phi_{[t,+\infty[}$ given by
$h\otimes g \mapsto f$ where $f(\sigma ) = h(\sigma \cap [0,t]) g
(\sigma  \cap (t,+\infty[)$. 

We shall use the following notations:
$$
\F_{t]}=\F_{[0,t]},\qq\F_{[t}=\F_{[t,+\infty[}.
$$
\smallskip
Define $\O$ to be the {\it vacuum vector}, that is,
$\O(\s)=\d_\emptyset(\s)$.
\bigskip
We now define a particular family of curves in $\F$, which is going to be of
great importance here. Define $\chi^i _t \,{\in}\, \Phi$ by
$$
\chi^i _t(\sigma ) = \cases {
\Unn_{[0,t]}(s) &if~~$\sigma = \{(s,i)\} $\cr 0 &otherwise. }
$$
Then  notice that for all $t\in\Rp$ we have that $\chi^i _t$ belongs
to $\Phi_{[0,t]}$. We actually have much more than that: we have 
$$\chi^i_t -
\chi^i_s \, \in \,  \Phi_{[s,t]} \hbox{\ for all\ } s\leq t.
$$
This last property can be checked immediately from the definitions, and
it is going to be of great importance in our construction. Also notice
that $\x^i_t$ and $\x^j_s$ are orthogonal elements of $\F$ as soon as $i\not
=j$. As  we will see later on, apart from trivialities, the curves
$\pro {\x^i}$ are the only ones to share these properties. 
\bigskip
These properties allow to define the so-called {\it Ito integral\/} on
$\Phi $. Indeed,  let $g=\{(g^i_t)_{t\ge
0}, \, i \in\L\}$ be families of elements of $\ \Phi $ indexed by both $\RR_+$ and $\L$, such that
\smallskip
 i) $t\mapsto \|g^i_t\|$ is measurable, for all $i$,
\smallskip
ii) $g^i_t \,{\in}\, \Phi _{[0,t]}$ for all $t$,
\smallskip
iii) $\sum_{i\in\L}\int^\infty _0 \|g^i_t\|^2\ dt < \infty $
\smallskip
\def\nb{\NN}
\noindent then one says that $g$ is {\it Ito integrable}
and we define its {\it Ito integral} 
$$
\sum_{i\in\L}\int^\infty _0 g^i_t\ d\chi^i_t
$$
 to be the
limit in $\Phi $ of
$$
\sum_{i\in\L} \sum_{j\in\NN} \wt g^i_{t_j}
\otimes \left(\chi^i _{t_{j+1}} - \chi^i_{t_j}\right)\eqno(2)
$$
where $\rS=\{t_j,~j{\in} \nb\}$ is a partition of $\ \rb^+$ which is understood to
be refining and to have its diameter tending to $0$, and $(\wt g^i_\cd)_i$
is an Ito integrable family in $\F$, such that for each $i$, $t\mapsto \wt g^i_t$ is a step process, and which converges to $(g^i_\cd)_i$ in
$L^2(\Rp\times\rP)$.

Note that by assumption we always  have that $\wt g^i_{t_j}$ belongs to
$\F_{t_j]}$ and $\x^i_{t_{j+1}}-\x^i_{t_j}$ belongs to $\F _{[t_j, t_{}j+1]}$, hence the tensor product symbol in (2). 

Also note that, as an example, one can take
$$
\wt g^i_t = {1\over t_{j+1}-t_j} \int^{t_{j+1}}_{t_j} P_{t_j} g^i_s\ ds
$$
where
$P_t$ is the orthogonal projection onto $\Phi _{[0,t]}$. 
\bigskip
One then obtains the following properties ({[At1], Proposition 1.4}):

\th{3.}{\it The Ito integral $I(g)=\sum_i\int^\infty _0 g^i_t \
d\chi^i_t$, of an Ito integrable family $g=(g^i_\cd)_{i\in\L}$,  is the  element of $\Phi $
given by
$$
I(g)(\s)=\cases{0& if $\s=\emptyset$\cr
g^i_{\vee\s}(\s\setminus(\vee \s,i))& if $\vee\s\in\s_i$.\cr}
$$
It satisfies the {\rm Ito isometry formula}:
$$
\normca{I(g)}=\Big\|{\sum_{i\in\L}\int^\infty _0 g_t^i \ d\chi^i _t} \Big\| ^2 = \sum_{i\in\L}\int^\infty _0
\norme{g^i_t}^2\,dt~.\eqno(3)
$$}
\qed
\bigskip
In particular, consider a family $f=(f^i)_{i\in\L}$ which belongs to $
L^2(\rP_1)=L^2(\Rp\times\L)$, 
then the family  $(f^i(t) \O)$, $t\in\Rp$, $i\in\L$, is clearly
Ito integrable. Computing its Ito integral we find that
$$
I(f)=\sum_{i\in\L}\int_0^\infty f^i(t)\O\, d\x^i_t
$$
is the element of the first particle space of the Fock space $\F$ associated to the
function $f$, that is,  
$$
I(f)(\s)=\cases{f^i(s)&if $\s=\{s\}_i$\cr
0&otherwise.\cr}
$$
\bigskip
Let us define the ``adjoint'' mapping of the Ito integral. For all
$f\in\F$, all $i$ in $\L$ and all $t\in\Rp$, consider the following
mapping on $\rP$:
$$
\left[D^i_tf\right](\s)=f(\s\cup \{(s,i)\})\indic_{\s\subset[0,s[}.
$$
We then have the following result ([At1],
Theorem 1.6).

\thl{4.}{[Fock space predictable representation property] }{\it
For all $f\in \F$, all $i \in \L$ and for almost all $t\in\Rp$, the
mapping $D^i_tf$ belongs to $\F=L^2(\rP)$. Furthermore, the family
$(D^i_\cd f)_i$ is always Ito integrable and we have the 
representation
$$
f=f(\emptyset)\O+\sum_{i\in\L} \int_0^\infty D^i_tf\,
d\x^i_t\eqno{(4)}
$$
with the isometry formula
$$
\normca f=\ab{f(\emptyset)}^2+\sum_{i\in\L}\int_0^\infty
\normca{D^i_tf}\, dt.\eqno{(5)}
$$}
\qed

As an immediate corollary we get the following.
\co{5.}{\it The representation (4) of $f$ is unique. In particular, if
$g\in\F$ is of the form
$$
g=c\,\O+\sum_{i\in\L}\int_0^\infty h^i_t\, d\x^i_t
$$
then for almost all $t$, all $i$ in $\L$,
$$
D^i_tg=h^i_t\,.
$$
}
\qed
\bigskip
Let $f {\in} L^2(\rP_n)$, one can easily define the {\it iterated Ito
integral\/} on $\ \Phi $:
$$
I_n(f) = \int_{\rP_n}f(\s)\, d\chi_\s
$$
by iterating the definition of the Ito integral:
$$
I_n(f)=\sum_{i_1,\ld,i_n \in \L}\int_0^\infty\int_0^{t_n}\ld\int_0^{t_2}
f^{i_1,\ld, i_n}(t_1,\ld,t_n)\O\,\,d\x^{i_1}_{t_1}\, \ld\,
d\x^{i_n}_{t_n}.
$$
We obtain this way an element of $\F$ which is actually the
representant of $f$ in the $n$-particle subspace of $\F$, that is
$$
[I_n(f)](\s)=\cases{f^{i_1,\ld,i_n}(t_1,\ld,t_n)&if
$\s=\{t_1\}_{i_1}\cup\ld\cup\{t_n\}_{i_n}$\cr0&otherwise.\cr}
$$

For any $f\in \rP$ we put
$$
\int_\rP f(\s)\, d\x_\s
$$
to denote the series of iterated Ito integrals
$$
f(\emptyset)\O+\sum_{n=1}^\infty\, \sum_{i_1,\ld,i_n \in\L}\int_0^\infty\int_0^{t_n}\ld\int_0^{t_2}
f^{i_1,\ld, i_n}(t_1,\ld,t_n)\O\, d\x^{i_1}_{t_1}\, \ld\,
d\x^{i_n}_{t_n}.
$$

We then have the following representation ([At1], Theorem 1.7).

\thl{6.}{[Fock space chaotic representation
property]}{\it Any element 
$f$ of $\Phi $ admits an {\rm 
abstract chaotic representation}
$$f = \int_{\pcc} f(\sigma )\ d\chi _\sigma \eqno{(6)}$$
satisfying the isometry formula
$$\|f\|^2 = \int_{\pcc} |f(\sigma )|^2\ d\sigma \eqno{(7)}.$$
This representation is unique.}

\qed

The above theorem is the exact expression of the heuristics we wanted
in order to describe the space 
$$
\F=\bigotimes_{\Rp}\rH.
$$
Indeed, we have, for each $t\in\Rp$,  a family of elementary
orthonormal elements $\O$, $d\x^i_t$, $i\in\L$
(a basis of $\rH$) whose (tensor) products
$d\x_\s$ form a continuous 
basis of $\F$ (formula (6)) and, even more, form an orthonormal
continuous basis (formula (7)).
\bigskip
The attentive reader will have noticed that the only property of the
curves $\x^i_\cd$ that we really used is the fact that
$\x^i_t-\x^i_s$ belongs to $\F_{[s,t]}$ for all $s\leq t$. One can naturally
wonder if there exists another such family, which will then allow
another Ito integral and furnish another continuous basis for $\F$ via
another chaotic expansion property. 

Of course there are obvious curves that can be obtained from
the $\x^i_\cd$: for any function $f$ on $\Rp$ and any  $g\in
L^2(\Rp\times\L)$ put 
$$
y_t=f(t)\O+\sum_i\int_0^tg^i(s)\O \,\, d\x^i_s
$$
for  all $t\in\Rp$. 
Then one easily
checks that  $(y_\cd)$ satisfies the same property, namely, $y_t-y_s$
belongs to $\F_{[s,t]}$ for all 
$s\leq t$. But clearly the Ito integration theory obtained from
$y$ is the same as the one from $\x$, except that scalar factors $g^i(s)$
will appear in the integration. 

One can wonder if there exist more complicated examples, giving rise
to a different Ito integration. The following result shows that there
are no more examples. In particular, there is only one Ito integral,
one chaotic expansion  and one natural continuous basis ({[At1], Theorem 1.8}). 

\th{7.}{\it Let $\pro y$ be a  curve in $\F$ such that
$y_t-y_s$ belongs to $\F_{[s,t]}$ for all $s\leq t$.
Then there exist a function  $f$ on $\Rp$ and $g\in
L^2(\Rp\times\L)$ such that 
$$
y_t=f(t)\O+\sum_i\int_0^tg^i(s)\O \,\,d\x^i_s
$$
for all $t\in\Rp$.} 

\qed

\sspa{III.2}{The quantum noises}

The space $\F$ we have constructed is the natural space for defining
quantum noises. These quantum noises are the natural, continuous-time,
extensions of the basis operators $a^i_j(n)$ we met in the atom chain
$\TF$. 

As indicated in the heuristic discussion above, we shall deal with a
family of infinitesimal operators $da^i_j(t)$ on $\F$ which act on the
continuous basis $d\x_\s$ in the same way as they discrete-time
counterparts $a^i_j(n)$ act on the $X_A$. The integrated version of the
above 
heuristic infinitesimal formulas easily gives an exact formula for the action
of the operators $a^i_j(t)$ on $\F$:
$$
\eqalignno{
[a^0_i(t)f](\sigma ) &= \sum_{s{\in} \sigma_i \atop s\leq t} f(\sigma \setminus \{s\}_i),\cr
[a^i_0(t)f](\sigma ) &= \int^t_0 f(\sigma \cup  \{s\}_i)\ ds,\cr
[a^i_j(t)f](\sigma ) &= \sum_{s{\in} \sigma_j \atop s\leq t}\
f(\sigma\setminus\{s\}_j\cup\{s\}_i )\cr
[a^0_0(t)f](\s)&=t\, f(\s)\cr
}
$$
for $i,j\not=0$.
\bigskip
All these operators, except $a^0_0(t)$, are unbounded. But 
note that a good common domain to all  these operators is
\def\dc{\rD}
$$\dc = \Big\{ f{\in} \Phi ~;~~\int_{\pcc} |\sigma | ~|f(\sigma )|^2\
d\sigma  < \infty  \Big\}~.$$
\bigskip

\def\es{\ec _{\mathcal S}}

\def\sup{\hbox{sup}}
This family of operators is characteristic and universal in a  sense
which is close to the one of the curves $\x^i_t$. Indeed, one can
easily check that in the decomposition of
$\F\simeq\F_{s]}\otimes\F_{[s,t]}\otimes\F_{[t}$, the operators
$a^i_j(t)-a^i_j(s)$ are all of the form
$$
I\otimes (a^i_j(t)-a^i_j(s))_{\vert\F_{[s,t]}}\otimes I.
$$
This property is fundamental for the definition of the quantum
stochastic integrals and, in the same way as for $(\x^i_\cd)$,  these operator
families are the only ones to share that property (cf [Coq]).

This property allows to consider Riemann sums:
$$
\sum_k H_{t_k}\left(a^i_j(t_{k+1})-a^i_j(t_k)\right)\eqno{(8)}
$$
where $\rS=\{0=t_0<t_1<\ld<t_k<\ld\}$ is a partition of $\Rp$, where $\pro H$ is
 a family of operators on $\F$ such that  
\smallskip
-- each $H_t$ is an operator of the form $H_t\otimes I$ in the tensor product space
$\F=\F_{t]}\otimes\F_{[t}$ (we say that $H_t$ is a {\it $t$-adapted
operator} and that $\pro H$ is an {\it adapted process of operators}), 
\smallskip
-- $\pro H$ is a {\it step process}, that is, it is constant on intervals:
$$
H_t=\sum_kH_{t_k}\indic_{[t_k,t_{k+1}]}(t),
$$
and where the operator product
$H_{t_k}\left(a^i_j(t_{k+1})-a^i_j(t_k)\right)$ is actually a tensor
product of operators 
$$
H_{t_k}\otimes\left(a^i_j(t_{k+1})-a^i_j(t_k)\right).
$$
Note that, in particular, the above ``product'' is commutative and
does not impose  any new domain constraint. 

The resulting operator associated to the Riemann sum (8) is denoted by
$$
\int_0^\infty H_s\, da^i_j(s).
$$
If we denote by $T$ the above operator and by $T_t$ the operator
$$
\int_0^t H_s\, da^i_j(s)=\int_0^\infty H_s\indic_{[0,t]}(s) \, da^i_j(s)
$$
we can then compute the action of $T$ on a ``good'' vector $f$ of its
domain and we obtain (cf [A-M] for more details)
$$
Tf=\sum_{k\in\L}\int_0^\infty T_tD^k_tf\, d\x^k_t+\int_0^\infty H_tD^i_tf\,
d\x^j_s\eqno{(9)}
$$
with the notations: $D^0_t=P_t$ and $d\x^0_
t=dt$.
For general operator processes (still adapted but not step process
anymore) and general $f$, it is equation (9) which is kept as a
definition for the domain and for the action of the operator
$$
T=\int_0^\infty H_s\, da^i_j(s).
$$
The maximal domain and the explicit action of the above operator can
be described but  is not worth developing  here. The interested
reader may refer to [At3], chapter 12 or to [A-L]. There are particular 
domains where the definition simplifies. The one we shall use here is
the case of coherent vectors. 

Indeed, if $\f$ is any element of $L^2(\Rp, \rH ')$, 
consider the associated coherent vector $\e(\f)$ in $\F$. That is, 
$$
[\e(\f)](\s)= \prod_i\prod_{s\in\s_i}\f_i(s).
$$
Put 
$\f_0(s)=1$ for all $s$.
If $\f$
is such
that 
$$
\int_0^t\ab{\f_j(s)}^{(1-\d_{0i})(2-\d_{0j})}\norme{H_s\e(\f)}^{2-\d_{0j}}\, ds<\infty
$$
then $\int_0^tH_s\, da^i_j(s)$ is well-defined on $\e(\f)$ with
$$
\ps{\e(\psi)}{\int_0^tH_s\,
da^i_j(s)\,\e(\f)}=\int_0^t\bar\psi_j(s)\f_i(s)\ps{\e(\psi)}{H_s\,\e(\f)}\, ds.
$$
for all $\psi\in L^2(\Rp,\rH')$. 

\sspa{III.3}{Embedding and approximation by the Toy Fock space}

We now describe the way the atom chain and its basic operators can be
realized as a subspace of the Fock space and a projection of the
quantum noises. The subspace associated to the atom chain is attached
to the choice of some partition of $\Rp$ in such a way that the
expected properties are satisfied:
\smallskip
-- the associated subspaces increase when the partition refines and they
  constitute an approximation of $\F$ when the diameter of the
  partition goes to 0,
\smallskip
-- the associated basic operators are restrictions of the others when
   the partition increases and they constitute an approximation of the
   quantum noises when the diameter of the partition goes to 0.
\bigskip
Note that this approximation has  deep interpretations in terms of
approximations of $n$-dimensional classical noises by extremal random
walks in $\RR^n$ whose jumps take $n+1$ different values. This aspect
is developed in [A-P].
\bigskip
\def\scc{\rS}
Let $\ \scc = \{0 = t_0 < t_1 < \cdots <t_n < \cdots \}$ be a partition of
$\ \rb^+$ and $\delta (\scc) = \sup_i |t_{i+1}-t_i|$ be the diameter
of $\ \scc$. For $\ \scc$ being fixed, define $\ \Phi _n = \Phi
_{[t_{n-1},t_{n}]}$, $n{\in} \NNE$. We clearly have that  $\Phi $ is
naturally isomorphic to the countable tensor product $ \otimes_{n{\in}
\NNE} \Phi _n$ (which is understood to be defined with respect to the
stabilizing sequence $(\O)_{n{\in} 
\nb}$).
\def\idelt#1{\int^{t_{#1+1}}_{t_{#1}}}

For all $n{\in} \nb^*$, define for $i,j\in\L$
$$
\eqalign{
X^i(n) &= {\chi^i _{t_{n}}-\chi^i _{t_{n-1}}\over \sqrt {t_{n}-t_{n-1}}} \in \Phi
_n~,\cr
a^i_0(n) &= {a^i_0(t_{n})-a^i_0({t_{n-1}})\over \sqrt{t_{n}-t_{n-1}}}\circ P_{1]},\cr
a^i_j(n) &=  P_{1]}\circ\left(a^i_j(t_{n})-a^i_j(t_{n-1})\right)\circ P_{1]},\cr
a^0_i(n) &= P_{1]} \circ{a^0_i(t_{n})-a^0_i({t_{n-1}})\over
\sqrt{t_{n}-t_{n-1}}}~,\cr
a^0_0(n) &= \qquad P_{0]},\cr
}
$$
where for $i=0,1$, $P_{i]}$ is the orthogonal projection onto $L^2(\pcc_{ i})$ and
where the above definitions are understood to be 
valid on $\ \Phi _n$ only, the corresponding operator acting as
the identity operator $I$ 
on the others $\Phi _m$'s. 
\smallskip
For every $A\in\rP=\rP_{\NNE,\L}$, define $X_A$ from the $X^i(n)$'s in
 the same way as for $\TF$:
$$
X_A=\O\otimes\ld\otimes\O\otimes X^{i_1}(n_1)\otimes
\O\otimes\ld\otimes\O\otimes X^{i_2}(n_2)\otimes\ld
$$
in $\otimes_{n\in\NNE}\rH_n$.
 
 Define $T\Phi (\scc)$ to
be the space of $f {\in} \Phi $ which are of the form
$$
f =  \sum_{A{\in} \rP}f(A) X_A
$$
(note that the condition $\|f\|^2 = \sum_{A{\in} \rP} |f(A)|^2 <
\infty $ is automatically satisfied).

The space $T\Phi (\scc)$ is thus clearly identifiable to the spin chain
$\TF$. 
The space $T\Phi (\scc)$ is a closed subspace of $\ \Phi
$.  We denote by
$P_\rS$ the operator of orthogonal
projection from $\ \Phi $ onto $T\Phi (\scc)$. One can prove for example
that the projection of an exponential vector is an ``exponential vector" of the embedded toy Fock space: indeed, a direct computation shows that for any $\phi$ in $L^2(\Rp, \rH ')$, 
$$ \big(P_\rS \, \e (\f)\big) (A) = \prod_i\prod_{n \in A_i}\ut_i(n)$$
where the function $\ut$ belongs to $l^2(\NNE, \rH ')$ and is defined by
$$ \ut _i (n) = \frac{1}{\sqrt{t_{n}-t_{n-1}}} \int_{t_{n-1}}^{t_{n}}
\f _i (s) \, ds.$$ We will denote by $e(\ut)$ such a discrete time
version of a coherent vector.
\smallskip 
We shall now check that the above operators $a^i_j(n)$ act on
$\TF(\rS)$ in the same way as the the basic operators of $\TF$. 

\prp{8.}{\it We have, for all $i,j\in\L$
$$\eqalign{
&\cases{a^i_0(n)\,X^j(n)=\d_{ij}\O\cr
a^i_0 \,\O =0}\cr
\noalign{\vskip3pt}
&\cases{a^i_j(n)\, X^k(n)=\d_{ik}X^j(n)\cr
a^i_j \,\O =0}\cr
\noalign{\vskip3pt}
&\cases{a^0_i(n)\, X^j(n)=0\cr
a^0_i(n)\, \O =X^i(n)}\cr
\noalign{\vskip3pt}
&\cases{a^0_0(n)\, X^k(n) = 0 \cr
a^i_j \,\O = \O.}\cr}
$$
}

\prf This is a direct  application of the definitions and 
computations using equation (9), cf [At2] for details. For example:
$$
\eqalignno{
a^i_0(n) X^j(n) &= {1\over t_{n}-t_{n-1}} \left(a^i_0({t_{n}})-a^i_0({t_{n-1}})\right)
\int^{t_{n}}_{t_{n-1}} \O\ d\chi^j _t\cr
&= {1\over t_{n}-t_{n-1}} \left[ \sum_{k\in\L}\,\int^{t_{n}}_{t_{n-1}}
\left(a^i_0({t})-a^i_0({t_{n-1}})\right) \O\ d\chi^k _t + \int^{t_{n}}_{t_{n-1}}
\d_{ij}\O\ dt\right]\cr
&= {1\over t_{n}-t_{n-1}} \left(0 + (t_{n}-t_{n-1})\d_{ij}\O\right) =
\d_{ij}\O.\cr 
}
$$
And so on for the other cases.\qed
\bigskip
Thus the action of the operators $a^i_j$ on the $X^i(n)$ is
exactly the same as the action of the corresponding operators on the spin chain
of section II; the operators $a^i_j(n)$ act on $T\Phi (\scc)$ exactly in the same way as the
corresponding operators do on $\TF$. We have completely embedded the toy
Fock space in the Fock space.
\bigskip
The action of operators $a^i_j (n)$ on discrete exponential vectors as defined above will be most useful in the
sequel. The following lemma is deduced immediately from Proposition 8.
\le{9.}{\it For any $\phi,\,\psi$ in $L^2(\RR_+,\rH)$ and for any $t_n$-adapted operator $H_n$ 
the bracket
$$ \left\langle e(\tilde\phi), \, H_n a^i_j(n+1) e(\tilde\psi) \right\rangle$$
is equal to
$$ \conj{\tilde\phi_j} (n+1) \tilde\psi_i (n+1) 
\left\langle e(\tilde\phi _{n]}),\, H_n e(\tilde\psi _{n]})\right\rangle \left\langle e(\tilde\phi _{[n+2}),\, e(\tilde\psi _{[n+2})\right\rangle.
$$}
This lemma is the basis for our future computations involving discrete-time quantum stochastic integrals (for more precise treatment of this subject see [Pa1] or [Pa3]).
\bigskip
We are now going to see that the Fock space $\Phi$ and its basic
operators $a^i_j(t)$, $i,j\in\L\cup\{0\}$ can be approached by the toy Fock
spaces $T \Phi(\scc)$ and their basic operators $a^i_j(n)$.

We are given a sequence $(\rS_n)_{n{\in} \NN}$ of partitions which are
getting finer and finer and whose diameter $\delta (\scc_n)$ tends to
$0$ when $n$ tends to $+\infty $. Let  $T
\Phi(n) = T \Phi(\rS_n)$ and $P_n= P_{\rS_n}$,  for all $n{\in} \NN$.

\th{10.}{\it

i) The orthogonal projectors $P_n$ strongly converge to the identity
operator I on $\F$. That is, any $f\in\F$ can be approached in $\F$ by
a sequence $\seq f$ such that $f_n\in\TF(n)$ for all $n\in\NN$. 
\smallskip
ii) If $\ \scc_n = \{0 = t^n_0 < t^n_1 < \cdots < t^n_k <
\cdots \}$, then for all $t {\in} \rb^+$, the operators
$\sum_{k;t^n_k\leq t} a^i_j(k)$, $\sum_{k;t^n_k\leq t}
\sqrt{t^n_{k}-t^n_{k-1}}\, a^i_0(k)$, $\sum_{k;t^n_k\leq t}
\sqrt{t^n_{k}-t^n_{k-1}}\, a^0_i(k)$, and $\sum_{k;t^n_k\leq t}
({t^n_{k}-t^n_{k-1}})\, a^0_0(k)$ converge strongly on $\dc$ to $a^i_j(t)$,
$a^i_0(t)$, $a^0_i(t)$ and $a^0_0(t)$ respectively.
}

\prf
 i) As the $\scc_n$ are refining then the  $(P_n)_n$ forms
 an increasing family of orthogonal projection in $\Phi$.
 Let $P_\infty  = \vee_n P_n$. Clearly,
for all $s\leq t$, we have that $\chi^i _t - \chi^i _s$ belongs to
 Ran$P_\infty$. But by the construction of the Ito integral
and by Theorem 6, we have that the $\chi^i _t-\chi^i _s$ generate
$\Phi$. Thus $P_\infty  = I$.

\medskip
ii)  Let us check the
case of $a^0_i$.  A direct computation shows that, for $f {\in} \dc$
$$\bigg[\sum_{k;t^n_k\leq t} \sqrt{t^n_{k} - t^n_{k-1}} a^0_i(k) f\bigg]
(\sigma ) = \sum_{k;t^n_k\leq t} \Unn_{|\sigma_i \cap [t^n_{k-1},t^n_{k}]|=1}
\sum_{s{\in} \sigma_i \cap [t^n_{k-1},t^n_{k}]} f(\sigma \setminus \{s\}_i).$$
 Put $t^n = \sup \big\{ t^n_k{\in} \scc_n~; t^n_k \leq t\big\}$. We have
$$
\di{
\Big\| \left(\sum_{k;t^n_k\leq t} \sqrt{t^n_{k} - t^n_{k-1}} a^0_i(k)  - a^0_i(t)\right) f
\Big\|^2\hf\cr
= \int_\pcc \Big|\sum_{k;t^n_k\leq t} \Unn_{|\sigma_i \cap
[t^n_{k-1},t^n_{k}]|=1} \sum_{s{\in} \sigma_i \cap [t^n_{k-1},t^n_{k}]}
f(\sigma \setminus \{s\}_i) - \sum_{s{\in} \sigma_i \cap [0,t]} f(\sigma \setminus \{s\}_i)
\Big|^2\ d\sigma \hf\cr
\leq 2 \int_{\pcc} \Big|\sum_{s{\in} \sigma \cap [t^n,t]} f(\sigma \setminus \{s\}_i)
\Big|^2\ d\sigma +  2 \int_{\pcc} \Big|\sum_{k;t^n_k\leq t}
\Unn_{|\sigma_i \cap[t^n_{k-1},t^n_{k}]|\geq 2}\hf\cr
\hf\times \sum_{s{\in} \sigma_i \cap [t^n_{k-1},t^n_{k}]} f(\sigma
\setminus \{s\}_i) 
\Big|^2\ d\sigma.}$$
\noindent For any fixed $\sigma $, the terms inside each of the integrals above
converge to $0$ when $n$ tends to $+\infty $. Furthermore we have, for
$n$ large enough,
$$\eqalign{
\int_{\pcc} \Big|\sum_{s{\in} \sigma \cap [t^n,t]} f(\sigma \setminus
\{s\}_i)\Big|^2\ d\sigma
&\leq \int_{\pcc} |\sigma |\sum_{s{\in} \sigma\atop s\leq t+1}
|f(\sigma \setminus 
\{s\}_i)|^2\ d\sigma \cr
&= \int^{t+1}_0 \int_{\pcc} (|\sigma |+1) |f(\sigma )|^2\ d\sigma \
ds\cr
&
\leq (t+1) \int_{\pcc} (|\sigma |+1) |f(\sigma )|^2\ d\sigma \cr}
$$
\noindent which is finite for $f {\in} \dc$;
$$\eqalign{
\int_{\pcc} \Big|\sum_{k;t^n_k\leq t} &\Unn_{|\sigma_i
\cap[t^n_{k-1},t^n_{k}]|\ge 2}  \sum_{s{\in} \sigma_i \cap [t^n_{k-1},t^n_{k}]}f(\sigma \setminus
\{s\}_i)\Big|^2\ d\sigma\cr
&\leq \int_{\pcc} \Big(\sum_{k;t^n_k\leq t} \Unn_{|\sigma_i
\cap[t^n_{k-1},t^n_{k}]|\geq 2}\  \Big| \sum_{s{\in} \sigma_i \cap
[t^n_{k-1},t^n_{k}]}f(\sigma \setminus 
\{s\}_i)\Big|\Big)^2\ d\sigma\cr
&\leq \int_{\pcc} \Big(\sum_{k;t^n_l\leq t}\ \sum_{s{\in} \sigma_i \cap
[t^n_{k-1},t^n_{k}]} |f(\sigma \setminus
\{s\}_i)|\Big)^2\ d\sigma\cr
&= \int_{\pcc} \Big(\sum_{s{\in} \sigma_i \atop s\leq t^n}  |f(\sigma \setminus
\{s\}_i)|\Big)^2\ d\sigma\cr
&= \int_{\pcc}|\sigma |  \sum_{s{\in} \sigma_i \atop s\leq t^n}  |f(\sigma \setminus
\{s\}_i)|^2\ d\sigma\cr
&\leq (t+1) \int_{\pcc} (|\sigma |+1) \big|f(\sigma )\big|^2\ d\sigma<\infty
\cr
}$$
in the same way as above. So we can apply Legesgue's theorem. This
proves the result.

The other cases are treated in the same way. See [At2] for details.\qed
\bigskip
We have fulfilled our duties: not only the space $\TF(\rS)$ recreates
$\TF$ and its basic operators as a subspace of $\F$ and a projection
of its quantum noises, but, when $\d(\rS)$ tends to 0, this realisation
constitutes an approximation of the space $\F$ and of its quantum noises.

\bigskip To any operator $H$ on
$\F$ we can associate the projected operator $P_\rS HP_\rS$ which
acts on the atom chain only and which approximates $H$ (if $H$ is
bounded for example). 

\def\es{P_\rS}
 We wish to compute the corresponding projections 
of the quantum stochastic integral operators. We reduce our
computations to the case where integrals are of 
the type $H=\int_0^\infty  H_t^\epsilon \,da^i_j(s)$, with $(i,j)\not=(0,0)$, 
and satisfy the following conditions {\bf (HS)}:
\smallskip
-- the operator $H$ is bounded and
\smallskip
-- the integrands $H^i_j(t)$ are bounded  for all $t$ and
   $t\mapsto\norme{H^i_j(t)}$ is square integrable if one of $i$ or $j$
   is zero, essentially bounded otherwise.
\smallskip
Even though they are rather restrictive, these hypotheses will suffice
for our neeeds.

The following result is a consequence of the theory and the computations developed in  [Pa1] and  [Pa2] (adapted here to the case of higher
multiplicity) and we do not reproduce the proof here. It is also stated in the following form in [Pa3], chapter 4.

\th{11.}{\it
Let $(i,j)\not =(0,0)$ be fixed. Let $H = \int H^i_j(t) da^i_j(t)$ be a quantum stochastic integral
on $\Phi$ that satisfies the assumptions {\bf (HS)}. Then $\es H
\es$ is an operator on $\TF$ of the form
$$
\sum_{k,l}\sum_n h^k_l(n) \, a^k_l(n+1)
$$
where the sum is over all couples $(k,l)$ in $(\L \cup \{ 0\})^2$
different from $(0,0)$ and is meaningful in the weak sense. The
operators $h^k_l$ are given by :  
\smallskip
-- if both $i$ and $j$ are nonzero,
$$
h^k_l(n)=\d_{ki}\d_{lj}\frac{1}{t_{n+1}-t_n}\es\int_{t_n}^{t_{n+1}}
P_{t_n}H^i_j(t)\, dt 
$$
\smallskip
-- if $i=0$,
$$
h^0_l(n)=\d_{lj}\frac{1}{\sqrt{t_{n+1}-t_n}}\es\int_{t_n}^{t_{n+1}}
P_{t_n}H^0_j(t)\, dt 
$$
and for all $k\not=0$,
$$
h^k_l(n)=\d_{lj}\frac{1}{t_{n+1}-t_n}\es\int_{t_n}^{t_{n+1}}
P_{t_n}H^0_j(t)\left(a^0_k(t)-a^0_k(t_n)\right)\, dt 
$$
\bigskip
-- if $j=0$,
$$
h^k_0(n)=\d_{ki}\frac{1}{\sqrt{t_{n+1}-t_n}}\es\int_{t_n}^{t_{n+1}}
P_{t_n}H^i_0(t)\, dt 
$$
and for all $l\not=0$,
$$
h^k_l(n)=\d_{ki}\frac{1}{t_{n+1}-t_n}\es\int_{t_n}^{t_{n+1}}
P_{t_n}\left(a^l_0(t)-a^l_0(t_n)\right)H^i_0(t)\, dt. 
$$
}
\qed

\sspa{III.4}{Quantum Langevin equations}

In this article what we call quantum Langevin equation is actually a
restricted version of what is usually understood in the literature
(cf [G-Z]); by this we mean that we here study the so-called quantum
stochastic differential equations as defined by Hudson and
Parthasarathy and heavily studied by further authors. 

This type of
quantum noise perturbation of the Schr\"odinger equation is exactly
the type of equation which we will get as the continuous limit of our
Hamiltonian description of repeated quantum interactions.
 \bigskip
The aim of quantum stochastic differential equations is to study
equations of the form
$$
dU_t=\sum_{i,j \in \L \cup \{0\}}L^i_jU_t\, da^i_j(t),\eqno{(10)}
$$
with initial condition $U_0=I$. The above equation has to be
understood as an integral equation 
$$
U_t=I+\int_0^t\sum_{i,j \in \L \cup \{0\}} L^i_jU_t\, da^i_j(t),
$$
for operators on $\rH_0\otimes\F$,
the operators $L^i_j$ being bounded operators on $\rH_0$ alone which
are ampliated to $\rH_0\otimes\F$.

The main motivation and application of that kind of equation is that
it gives an account of the interaction of the small system $\rH_0$
with the bath $\F$ in terms of quantum noise perturbation of a
Schr\"odinger-like equation. Indeed, the first term of the equation 
$$
dU_t=L^0_0U_t\, dt+\ld
$$ 
describes the induced dynamics on the small system, all the other
terms are quantum noises terms.

One of the main application of equations such as (10) is that they give
explicit constructions of unitary dilations of semigroups of
completely positive maps of 
$\rB(\rH_0)$ (see [H-P]).  Let us
here only recall one of  the main existence, uniqueness  and boundedness
theorems connected to equations of the form (10). The literature is
huge about those equations; we refer to [Par] for the result we
mention here.

\th{12.}{\it If $\rH_0$ is separable and 
$$ \norme L = \big(\sum_{i,j \in \L\cup\{0\}} \normeca{L^i_j} \big) ^{1/2} \, < \, +\infty,$$
then the quantum
stochastic differential equation
$$
U_t=I+\sum_{i,j}\int_0^tL^i_jU_t\, da^i_j(t)
$$
admits a unique solution defined on the space of coherent vectors.

 The solution $\pro U$ is made of unitary operators if and only if
 there exist, on $\rH_0$, a self-adjoint operator $H$, operators $L_i$, $i\in\L$ and operators $S^i_j$, $i,j \in \L$ such that $(S^i_j)_{i,j \in \L}$ is unitary and the coefficients $L^i_j$ are of the form
$$
\eq{
L^0_0&=-(iH+\frac 12\sum_{k\in\L}L^\ast_kL_k)\cr
L^0_j&=L_j \vphantom{\sum_{k\in\L}}\cr
L^i_0&=-\sum_{k\in\L}L^\ast_kS^k_i\cr
L^i_j&=S^i_j-\d_{ij}I.\cr
}
$$
}\qed

\spa{IV}{Convergence theorems}

\sspa{IV.1}{Convergence to quantum Langevin equations}
We are now ready to assemble together all the pieces of the puzzle and
prove that the Hamiltonian dynamic associated to repeated quantum
interactions spontaneously converges to a quantum Langevin equation
under some normalization conditions on the Hamiltonian. Notice that we
no longer assume that $\LL(h)$ has been conveniently constructed for
our needs; in particular $\LL$ is not assumed to be unitary.

\def\R{\RR}

\def\LdR{L^2\left( \RR _+ \right)}

\def\R{\RR} 
\def\N{\NN}

\def\L{\LL}
\def\La{\Lambda}

\def\oepsr{\omega^\epsilon}
 
Let $h$ be a parameter in $\Rp$, which is thought of as representing a small time interval.
Let $\LL(h)$ be an operator on $\rH_0\otimes\rH$, with coefficients
$\LL^i_j(h)$ as a matrix of operators on $\rH_0$. Let $u_n(h)$ be
the associated solution of 
$$
u_{n+1}(h)=\LL_{n+1}(h)u_{n}(h)
$$
with the same notation as in section II.3.
In the following we will drop dependency in $h$ and write simply $\LL$, 
or $u_n$. Besides, we denote
$$
\e_{ij}=\frac 12(\d_{0i}+\d_{0j})
$$
for all $i,j$ in $\La\cup\{0\}$. That is, 
$$
\e_{i0}=e_{0j}=\frac 12,\ \ \e_{ij}=0, \ \ \e_{00}=1.
$$

Note that from now on we take the embedding of $\TF$ in $\F$ for
granted and we consider, without mentionning it,  all the repeated
quantum interactions to 
happen in $\TF(h)$, the subspace of $\F$ associated to the partition
$\rS=\{t_i=ih; i\in\NN\}$. We also make the convention that the
default summation sets for sums is $\L \cup \{0\}$, {\it e.g.} $\sum
_i$ is $\sum_{i\in\L \cup \{0\}}$.

\th{13.}{\it Assume that there exist bounded operators $L^i_j$, $i,j \in \La \cup \{0\}$ on
$\rH _0$ such that
$$\somu {i,j} \normeca{L^i_j}<+\infty$$
and
$$\lim_{h\rightarrow 0}
\somu {i,j}\normeca{\frac{\LL^i_j(h)-\d_{ij}I}{h^{\e_{ij}}} -L^i_j} = 0.
$$
for all $i,j=0,\ld,N-1$. Assume that the quantum stochastic
differential equation
$$
dU_t=\sum_{i,j}L^i_jU_t\, da^i_j(t)
$$
with initial condition $U_0=I$ admits a unique solution $\pro U$ which
is a process of bounded operators with locally uniform norm bound.

Then, for almost all $t$, for every $\f,\, \psi$ in $L^\infty([0,t])$,
the quantity 
$$
\ps{ a \otimes \e(\f) }{\es u_{[t/h]} \es\, b \otimes \e(\psi)}
$$   converges to $$\ps{  a \otimes \e(\f)}{U_t\, b \otimes \e(\psi)}$$
when $h$ goes to 0. 

Moreover, the convergence is uniform for $a$, $b$ in any bounded ball of $\rH$,  
uniform for $t$ in a bounded interval of $\R _+$.
}

\bigskip
\noindent{\bf Remarks}

\noindent -- This is where we particularize the index zero : the above hypotheses of convergence simply mean that, among the coefficients of $\LL$,
\par
\qq\qq $(\LL^0_0(h)-I) / h$ converges,\par
\qq\qq \ $\LL^i_j(h) / \sqrt h$ converges if either $i$ or $j$ is zero, $\ecarte$ \par
\qq\qq \ $\LL^i_j(h) - \delta_{i,j}$ converges if neither $i$ nor $j$
is zero. $\ecarte$
\smallskip
\noindent We here meet the announced three time scales appearing in the
Hamiltonian. We shall discuss the physical meaning of these
normalizations in next section.
\smallskip
\noindent -- In the case where the operator $\LL$ is unitary and satisfies the convergence assumptions of the above theorem, then one can see that the limiting operators $L^i_j$ are of the form given in the second part of Theorem 11.
\smallskip
\noindent -- In that case, the solution $(U_t)_{t\in\RR _+}$ enjoys a particular algebraic property which we won't define here: it is a {\it cocycle} (see [H-P] or chapter 6 of [Pa3]). This property traduces the fact that the evolution of the system is, in the limit, memory-less.
\bigskip

\def\oepsr{\omega^\epsilon}

\def\sd{\sqrt h}
\def\d{\delta}
Consider the  quantum stochastic differential equation $(E)$ on 
$\rH_0 \otimes \Phi$:
$$
 dU_s = \sum_{i,j}L^i_jU_s\, da^i_j(s)
$$
where the $L^i_j$ are the bounded operators on the initial space
$\rH_0$ given by our assumptions.
 
We consider that $h$ is fixed and the associated partition
$\rS=\{0=t_0<t_1=h<\ld<t_k=kh<\ld\}$ is also fixed. Note that we have
chosen a regular partition only for simplicity and that all our
results hold with general partitions when the mesh size tends to 0. We 
fix some bounded interval $[0,T]$ of $\R _+$.
 
We will proceed by successive simplifications.  
Consider the operator on the atom chain defined by $$w_k = \es U_{t_k}
\es.$$
The following lemma will be used over and again.

\le{14.}{\it 
For any $r<s$, any vectors $a \otimes \e(\f)$, $b \otimes \e(\psi)$
with $$\f,\psi\in L^2(\Rp;\rH') \cap L ^\infty(\Rp;\rH'),$$ we have 
$$
\ab{\ps{a \otimes \e(\f)}{(U_s - U_r) \, b \otimes \e(\psi)}} 
\leq C \norme a \norme b
(s-r)
$$ 
where $C$ depends only on $\norme L$, defined in Theorem 12,
and on the $L^2$ and $L^\infty$ norms of $\f$ and $\psi$. 
}

\prf
$$
\di{
\ \ab{\ps{a \otimes \e(\f)}{(U_s - U_r) b \otimes \e(\psi)}} 
 \hf\cr 
\hf\leq \sum_{i,j}\int_r^s\ab{\bar\f_i(u)}\,\ab{\psi_j(u)}\, \ab{\ps{a
\otimes \e(\f)}{L^i_jU_u b \otimes \e(\psi)}}\, du\cr
\hf\leq \norme L \int _r^s \norme{\phi(u)}\norme{\psi(u)} \norme{a \otimes \e(\phi)} \norme{U_u b \otimes \e(\psi)}\,du \hp{\sum_{i,j} L}} 
$$ 
from which the estimate follows, using the fact that $\norme{\f(u)}\leq
\norme\f_\infty$ and $\norme{\psi(u)}\leq
\norme\psi_\infty$, the $L^i_j$ are bounded and $U$ is locally
uniformly bounded.\qed
\bigskip
The following lemma shows that ${(w_k)}_{k}$ converges to $\pro U$, in
the same weak sense as in the theorem, as $h$ goes to 0.

\le{15.}{\it 
For any $t_k<s$, any vectors $a \otimes \e (\f)$, $b \otimes \e
(\psi)$ with $$\f,\psi\in L^2(\Rp) \cap L ^\infty(\Rp),$$ we have 
$$
\di{
\ab{\ps{a \otimes \e(\f)}{(w_k - U_s) b \otimes \e(\psi)}} 
\hf\cr
\hf\leq C \norme a \norme b
\left((s-t_k)+\norme{(I-\es)\e(\f)}+\norme{(I-\es)\e(\psi)}\right).\cr
}
$$ 
where $C$ depends only on $\norme L$
and on the $L^2$ and $L^\infty$ norms of $\f$ and $\psi$. 
} 
\def\conj{\bar}
\prf 
We have  
$$
\eq{
\vert<
{a\otimes \e(\f)}\, ,\,(w_k -& U_s)  b \otimes
\e(\psi) >\vert
\leq\cr
&\leq\ab{\ps{ a \otimes \e(\f)}{(U_{t_k} - U_s)  b
\otimes \e(\psi) }}\cr
& \ \ +\ab{\ps{ a \otimes
\e(\f)}{(\es U_{t_k}\es - U_{t_k}\es) b \otimes \e(\psi)}}\cr
&\ \ + \ab {\ps{a \otimes
\e(\f)}{( U_{t_k}\es - U_{t_k})b \otimes \e(\psi) }}\cr
&\leq\sum_{i,j}\int_{t_k}^s\ab{\bar\f_i(u)}\ab{\psi_j(u)}\ab{\ps{a\otimes\e(\f)}{L^i_j
U_ub\otimes\e(\psi)}}\, du
\cr
&\ \ +\norme{(I-\es)a\otimes\e(\f)}\norme{U_{t_k}\es
b\otimes\e(\psi)}\cr
&\ \ 
+ \norme{ U_{t_k}^\ast a\otimes\e(\f)}\norme{(I-\es) b \otimes
\e(\psi)}\cr
}
$$
and we conclude as in the previous lemma.
\qed
\vfill \eject
\noindent We can now prove Theorem 13.
\smallskip
\noindent{\bf Proof of Theorem 13}

Let $\o^i_j(h)$ be such that 
$$
\LL^i_j(h)-\d_{ij}I=h^{\e_{ij}}(L^i_j+\o^i_j(h))
$$
for all $i,j$ in $\La \cup \{0\}$. In particular
we have
that, $$ \sum _{i,j \in \La \cup \{0\}} \normeca{\o^i_j(h)}$$ converges to 0 when $h$ tends to 0. 

Consider the
solution $\seqe u$ of
$$
u_{n+1}=\LL_{n+1}u_n
$$
 with the notations of section II.3. Note that if $A$ denotes the matrix
 $\LL-(\d_{ij}I)_{i,j}$ we then have
$$
u_{n+1}-u_n=A_{n+1}u_n.
$$
Let $F$ be the matrix $(h^{\e_{ij}}L^i_j+\wh\d_{ij}hL^0_0)_{i,j}$ where 
$$
\wh\d_{ij}=\cases{1 & if $i=j$ and $(i,j)\not=(0,0)$,\cr 0&if
$i\not=j$ or $(i,j)=(0,0)$\cr}
$$
 and consider the solution $\seq
v$ of the
equation
$$
v_{n+1}-v_n=F_{n+1}v_n.
$$
Note that
$$
A_{n+1}=\sum_{i,j}A^i_ja^i_j(n+1)
$$
and similarly for $F_{n+1}$.

Also note that $a^i_j(n+1)$ commutes with $u_n$ ({\it resp.} $v_n$), for they do not act on the
same part of the space $\TF$.
Thus we get another useful way to write the above equations in terms of the
basis $a^i_j(n)$:
$$
u_{n+1}-u_n=\sum_{i,j}A^i_ju_n\, a^i_j(n+1).
$$
and
$$
v_{n+1}-v_n=\sum_{i,j}\left(h^{\e_{ij}}L^i_j+\wh\d_{ij}hL^0_0\right)v_n\,
a^i_j(n+1). 
$$ 
From the above lemma
it is enough to prove the convergence to zero of $u_n - w_n$. We
actually start with  
$w_n - v_n$. 
 
From the fact that 
$$
U_{t_{k+1}} - U_{t_k} = \sum_{i,j}\intk k L^i_j U_s\, da^i_j(s)
$$  
and thanks to the formulas for projections of Fock space integrals onto the toy Fock space in Theorem 11, 
one obtains the following expression for $w_{k+1} - w_k$ (be careful
that the $da^0_0(t)$ integrals give rise to $a^i_i(k)$ terms for {\it
all} $i$, for $I=\sum_i a^i_i$):
$$
\eq{ 
w_{k+1}& - w_k=  
 \sum_{i,j\not=(0,0)} h^{\e_{ij}}L^i_j \left( \frac{1}{h} \, \es
 \intk k P_{t_k} U_t \,dt\right)\, a^i_j(k+1) \es \cr 
& \qq + \sum_{i}\ h L^0_0 \left(\frac1h \es \intk k U_t \, dt  \right)
 a^i_i (k+1)\es 
\cr 
 & \qq+ \sum_{i\in\La}\sum_{j\in\La}\es \left(  \frac{1}{h} \intk k
\left(L^i_0 P_{t_k} 
U_t(a^j_0(t)-a^j_0(t_k))\right) \,dt\right.\cr
& \qq\left.+  \frac{1}{h} \intk k\left(L^0_j P_{t_k}
U_t (a^0_i(t)-a^0_i(t_k))\right) \,dt \right)  
a^i_j(k+1) \es.\cr}
$$
As a consequence 
$$
\eq{
w_n - v_n  
 &= \somu {k<n}\  
  \sum_{(i,j)\not=(0,0)}h^{\e_{ij}}L^i_j\, a^i_j(k+1) \left(
 \frac{1}{h} \, \es  \intk k P_{t_k} U_t \,dt\es-v_k\right) 
\cr 
& \qq + \somu{k<n} \, \sum_i h L^0_0 \, a^i_i(k+1) \left( \frac1h \es
\intk k U_t \, dt \es - v_k 
\right) \cr
 & \qq+\somu {k<n} \sum_{i\in\La}\sum_{j\in\La}\es \left(  \frac{1}{h}
 \intk k 
\left(L^i_0P_{t_k} U_t (a^j_0(t)-a^j_0(t_k))\right) \,dt\right.\cr
& \qq\left.+  \frac{1}{h} \intk k \left(L^0_j P_{t_k}
U_t(a^0_i(t)-a^0_i(t_k))\right) \,dt \right)\es \, 
a^i_j(k+1) .&(11)\cr}
$$
  We first wish to replace  $\frac{1}{h} \, \es \intk k P_{t_k} U_t \, dt $  
or $\frac1h \, \es \intk k U_t\,dt$
in the first two terms by $w_k$. 
 Lemma 14 allows us to estimate the error term. Consider two
essentially bounded   
functions $\f$, $\psi$ in  
$\LdR$ and two vectors $a, \, b$ in the unit ball ${\cal H} _1$ of $\rH$; 
we expand the first terms in (11) and estimate each term : 
$$
\di{ 
\ab{\ps{ a \otimes
e(\ut)}{\somu {k<n} h^{\e_{ij}}L^i_j\left( \frac{1}{h} \es 
\intk k\, P_{t_k} U_t \,dt - w_k \right) a^i_j(k+1) \, b \otimes e(\vt)}} \hf\cr  
   \hf\leq \somu {k<n} \ab{\ut_j (k)}\ab{\vt_i (k)} \ab{\ps{  a  
\otimes e(\ut _k)}{\es  \, h^{\e_{ij}} L^i_j 
\frac{1}{h}\!\! \intk k  (U_s  - U_{t_k}) ds\, b \otimes e(\vt _k) }},\cr}
$$
where  we have omitted a uniformly bounded factor 
$$
\left\langle a\otimes e(\ut _{[k+2}), b \otimes e({\vt
_{[k+2}})\right\rangle.
$$
From the fact that $\sum_i \ab{\ut_i(k)} \leq \sd \normei \f$, $\sum_i\ab{\vt_i (k)} \leq 
\sd \normei \psi$ we obtain an estimation of the error term of the form 
$$C\norme a \norme b \, \sqrt h $$ 
for some constant $C$ which depends only on the $\norme L$  and on the  
$L^2$ and $L^\infty$ norms of $\f$ and $\psi$. 
 
On the other hand, the error term obtained by replacing  $\frac1h \, \es \intk k U_t\,dt$
by $w_k$ in (11) is clearly dominated by $C\,h$ in norm because
$\sum_i h L^0_0 a^i_i (k+1)$ is just $h L^0_0$.

We now seek to evaluate the third sum in (11); for that  
consider again two functions $\f,\, \psi$ in $L^\infty ([0,t])$ and two vectors $a, \, b$ in the unit ball
${\cal H}_1$ of $\rH$; we have, up to a uniformly bounded factor,
$$ 
\di{ \sum_{i,j\in\La}\! \ab{\ps{ a \otimes \e(\f) }{ \frac{1}{h}L^i_0 
\es \Big(\intk k \!\! P_{t_k} U_s (a^j_0( s) - a^j_0(t_k)) \,ds\Big)
\, a^i_j(k+1) \, \es\, b 
\otimes \e (\psi)}}\hf\cr
\ =  
\sum_{i,j\in\La} \ab{\ut_j (k) \vt_i(k)
} 
\ab{
\ps{ a\otimes e (\wt \f _{k}) }{\frac 1h L^i_0\idelt k  U_s  (a^j_0( s) - a^j_0(t_k)) \, b \otimes e (\wt\psi _{k})\, ds}}\hf\cr
\ \leq C h ^{3/2} \norme{L} \normei \f \normei \psi \ecarte,\hf\cr
}
$$ 
for
 $a^j_0( s) - a^j_0(t_k)$ is bounded on $\Phi _{t_k}$, with norm $\sqrt 
{s - t_k}$. One obtains similarly
$$
\di{\sum_{i,j\in\La}
\ab{
\ps{ a \otimes \e(\f) }{\frac{1}{h} L^0_j \, \es \intk k
P_{t_k} U_s(a^0_i( s) - a^0_i(t_k))  \,ds \, a^i_j( k+1) \, \es b
\otimes \e (\psi)}
}\hf\cr
\leq  h ^{3/2} \norme{L} \normei \f \normei \psi ,\hf\cr
}
$$
so that the third sum in (11) is bounded by $C\, \sqrt{h}\,  2t  \normei \f \normei \psi$. 
\bigskip

We have shown that, putting $F_k=\sum_{i,j}(h^{\e_{ij}}L^i_j + \wh
\d_{ij}hL^0_0)\, a^i_j(k+1)$, 
$$
\di{
\ab{ \ps{a \otimes \e (\f)}{\es (w_n - v_n) \es b \otimes \e (\psi)} } \hf \cr
\hf =  \ab{\somu {k<n} \ps{  a \otimes e (\ut)}{ F_k (w_k - v_k) \, b \otimes e
(\vt)}} + o(1) \qq (12)}$$
where $o(1)$ is a term which
converges to zero as $h$ goes to zero {\it uniformly for $a$, $b$ in ${\cal H} _1$ and
for $t$ in a bounded interval}. That
uniform convergence property will be  important in the sequel. 
 
Expanding $F_k$ in equation (12) gives 
$$
\di{
  \ab{\ps{  a \otimes \e ({\ut _n})}{(w_n - v_n) \es \, b \otimes \e ({\vt _n})}}
\hf \cr
\qq\qq\leq   \sum_{i,j}\somu {k<n} h^{\e_{ij}}\ab {\ut_j (k)} \ab
{\vt_i (k)} \ab{\ps{ (L^i_j)^\ast a \otimes e(\ut _k)}{ (w_k - v_k)  b
\otimes e (\vt _k) }}\hf\cr 
\qq\qq \ + \sum_{i\in\La} \somu {k<n} h \ab {\ut_i (k)} \ab {\vt_i
(k)} \ab{\ps{ (L^0_0)^\ast a \otimes e(\ut _k)}{ (w_k - v_k)  b \otimes
e (\vt _k) }}\hf\cr
\qq\qq\  + o(1),\hf\cr}
$$
where in each term we have omitted an uniformly bounded factor
and the notation 
$o(1)$ indicates a function of $h$ which converges to zero as $h$ goes to zero.
 
Since $\f$ and $\psi$ are essentially bounded, the quantities
$\sum_{i,j} h^{\e_{ij}} \ab{\ut_j (k)} \, \ab{\vt_i(k)}$ and
$\sum_{i\in\La} h \ab{\ut_i (k)} \, \ab{\vt_i(k)}$ are again of order   
$h$ at most, with an estimate which is independent on $k$. Besides, remark that normalizing all operators would only imply
an additional constant factor, so that we can assume all the $L^i_j$ to be contractions. In that case the above implies that
for all $n$,  
$$
\di{
\sup _{a,b \, \in \, {\cal H} _1}  
\ab{\ps{ a \otimes e (\ut _n)}{ (w_n - v_n) b \otimes e (\vt _n) }}\hf\cr 
\hf\leq h C \somu {k<n} \sup _{a,b  \, \in \, {\cal H} _1}
\ab{\ps{  a \otimes e(\ut _k)}{(w_k - v_k) b \otimes e(\vt _k)}} 
+ o(1) \cr}
$$  
for some constant $C$. Here we have used our earlier remark that all
convergences are  
uniform in $a,\, b \, \in {\cal H} _1$. 
This implies that 
$$ \sup _{a,b  \, \in \, {\cal H} _1}  
\ab{\ps{ a \otimes e (\ut _n) }{ (w_n - v_n) b \otimes e (\vt _n)} }
\leq   (1+ C h)^n \times o(1) $$ 
and since $n h$ converges to t, the quantity   
$(1+ C h)^n$ is bounded so that 
$$
\sup _{a,b \, \in \, {\cal H} _1}  
\ab{\ps{ a \otimes e (\ut _n) }{(w_n - v_n) b \otimes e (\vt _n) }}
$$ 
converges to zero as $h$ goes to zero. 
 
We have proved the desired convergence property for the process $(w_k)_{k\geq0}$. 
Now we will prove that  
$$
\sup _{a,b \, \in \, {\cal H} _1}  
\ab{\ps{  a \otimes e (\ut _n) }{(u_n - v_n) b \otimes e (\vt _n) }}
$$ 
converges to zero as $h$ goes to zero. We have
$$
\eq{  
u_n - v_n &= \somu {k<n} F_k (u_k - v_k) + \somu {k<n} \left(
\sum_{i,j}h^{\e_{ij}}(\o^i_j(h) -h\wh\d_{ij}L^0_0) a^i_j(k+1)\right) u_k 
\cr
&= \somu {k<n} \left(F_k +
\sum_{i,j}h^{\e_{ij}}(\o^i_j(h)-h\wh\d_{ij}L^0_0)a^i_j(k+1)\right)  (u_k - v_k) \cr &\ \   +
\somu {k<n}\left(\sum_{i,j}h^{\e_{ij}}(\o^i_j(h)-h\wh\d_{ij}L^0_0)a^i_j(k+1) \right)
v_k\hf\qq\qq
\qq\qq\qq\qq{(13)}
\cr
}
$$
Replace for simplicity  $\o^i_j(h)$ by $\o^i_j(h)-h\wh\d_{ij}L^0_0$ (which does not change the validity of the earlier estimates).
The interest of the second form lies therein, that the term without
recurring $u_k - v_k$  
can now be estimated thanks to our previous result:  
$$
\ab{\ps{  a \otimes
e(\ut _k) }{\somu {k<n}
\left(\sum_{i,j}h^{\e_{ij}}\o^i_j(h)a^i_j(k+1)\right) v_k
 \, b \otimes e(\vt_k)}}  
$$   
is bounded by 
$$
\somu {k<n} \left(\sum_{i,j}h^{\e_{ij}}\ab{\ut_j (k)}\ab{\vt_i(k)}
\norme{\o^i_j(h)} \right)   \, \sup _{a,b \, \in \, {\cal H}_1}
\ab{\ps{ a \otimes e (\ut _k)}{ v_k b \otimes e(\vt _k )} } \eqno{(14)}
$$  
with 
$$
\di{ 
  \sup _{a,b \, \in \, {\cal H} _1} \ab{\ps{
         a \otimes e (\ut _k)}{ v_k b \otimes e(\vt _k )} } \hf\cr  
        \qq\leq \sup _{a,b \, \in \, {\cal H} _1} \left( \ab{\ps{ a \otimes e (\ut _k)}{(v_k
- w_k) b \otimes e(\vt _k )} }\right. \hf\cr
       \hf+\left.\ab{\ps{ a \otimes e (\ut _k)}{ w_k b \otimes e(\vt _k )} }\right) 
\cr}
$$
and our estimate  shows that the first term on the
right-hand-side  converges to zero uniformly in $k$ as $h$ goes to
zero. The second term is, in turn, bounded by $\norme {\e (\f)}
\norme{\e (\psi)}$ since any $w_k$ is $\es U_{t_k} \es$ and as such has bounded norm.

Thanks to our assumptions on the perturbative operators, the bound (14)
we are interested in converges to zero as $h$ goes to zero, uniformly for $a$, $b$  
in ${\cal H} _1$. 
 
Besides, for the recurring term in  (13) we obtain as before 
$$ 
\di{
 \sup _{a,b \in{\cal H}_1} \! \bigg|{\ps{ a \otimes e (\ut _n)}{\somu {k<n}
\Big(F_k + \sum_{i,j}h^{\e_{ij}}\o^i_j(h)a^i_j(k+1)\Big) (w_k
- v_k) b \otimes e(\vt _n) }}\bigg| \cr 
\hf\leq h \, C\somu {k<n} \sup _{a,b \, \in \, {\cal H} _1} \ab{\ps{  a \otimes e (\ut _k)}{(v_k -
w_k) b \otimes e(\vt _k) } }, \cr
} 
$$ 
thanks to the fact that the operators $\oepsr$ are assumed to have norms  
which converge to zero uniformly with $h$. We conclude as in the previous case. 

This ends the proof. 
\qed 

 \bigskip
Under some additional assumptions, which are verified in many
applications, we can very much improve the convergence.

\th{16.}{\it Consider the same assumptions and the same notations as in
Theorem 13. If furthermore $\norme{u_k}$ is locally uniformly bounded,
then $u_{[t/h]}$ converges weakly to $U_t$ on all $\rH_0\otimes\F$.}

\prf
Theorem 13 allows us to perform a $\epsilon / 3$ argument with an 
approximation of any vectors of $\rH_0\otimes \TF$  by combinations of vectors
$a \otimes \e (\f)$,  $b \otimes \e (\psi)$ with essentially bounded functions 
$\f$, $\psi$. 
\qed
\bigskip
One of the main application of this last theorem is  the case when the matrices $\LL(h)$ give rise at the
limit to a matrix $L$ such as in Theorem 12 (the case of a unitary
solution $\pro U$). We shall show that the associated discrete
evolution $\seq u$ satisfy the conditions of the above theorem, so
that the  convergence of $u_{[t/h]}$ towards $U_t$ is weak.

\th{17.}{\it Consider a matrix $\LL$ on $\rH_0\otimes \rH$ with
coefficients 
$$
\eq{
\LL^0_0&=I-h(iH+\frac 12\sum_kL^\ast_kL_k)+h\o^0_0\cr
\LL^0_j&=\sqrt hL_j+h\o^0_j\cr
\LL^i_0&=-\sqrt h\sum_kL^\ast_kS^k_j+h\o^i_0\cr
\LL^i_j&=I+S^i_j-\d_{ij}I+h\o^i_j\cr
}
$$
where  $H$ is a bounded self-adjoint operator, $(S^i_j)_{i,j \in \La}$,
is unitary, the $L_i$, $i\in\La$ are operators on $\rH_0$ such that $\sum L_k^* L_k$ converges strongly, the coefficients $\o^i_j$ are such that $\somu{i,j}\normeca{\o^i_j}$ is uniformly bounded and $\norme \o^0_0(h)$ converges to 0 when $h$ tends to~0.

Then the solution $\seq u$ of 
$$
u_{n+1}=\LL_{n+1}u_n
$$
is made of invertible operators which are locally uniformly bounded
in norm.  

In particular $u_{[t/h]}$ converges weakly to the solution $U_t$ of
the quantum stochastic differential equation $(E)$.
}

\prf

A straightforward computation shows the special form of $\LL$ induces
many cancellations when computing  the coefficients of
$\LL^\ast\LL-I$ and of $\LL\LL^\ast-I$, and that they are of order $h$. Thus for $h$ small enough the
operators $\LL^\ast\LL$, $\LL\LL^\ast$ and thus $\LL$ are
invertible. Thus so are the operators $u_n$. 

Furthermore the above estimates show that
$\norme{\LL}\leq\sqrt{1+Ch}$. This easily gives the locally uniform
boundedness of $\seq u$ and thus the desired weak convergence.\qed
\smallskip

Specializing some more will allow us to answer the natural question of
convergence of Heisenberg evolutions of observables:
\co{18.}{\it If the operator $\L$ is unitary and satisfies the conditions
of Theorem 13 then the solution $(U_t)_{t\in\R_+}$ of (E) is
unitary. In this case the convergence of $u_{[t/h]}$ to
$U_t$ is strong and for all bounded operator $X$ on $\rH_0 \otimes
\Phi$, almost all $t$, the sequence $u_{[t/h]}^* X u_{[t/h]}$
converges weakly to $U_t^* X U_t$ as $h$ tends to zero.}

\prf
It is easy to see from the above conditions that the operators $(L^i_j)_{i,j}$
satisfy for all $i,j$
$$ L^i_j + L^i_j {}^* + \sum_{k\in \La} L^i _k L^j_k {}^* = 0$$
$$ L^i_j + L^i_j {}^* + \sum_{k\in \La} L^k _i L^k_j {}^*  = 0$$
which implies that the equation (E) is of the form which has unitary
solutions (see Theorem 12).

By Theorem 17 and Proposition 1, for almost all $t$, $u_{t/h}$ is a
sequence of unitary operators that converges weakly to a unitary
operator, so that strong convergence also holds.

It is now straightforward to prove the statement regarding convergence
of $u_{[t/h]}^* X u_{[t/h]}$ to $U_t^* X U_t$ for bounded $X$.
\qed

\def\L{\Lambda}
\sspa{IV.2}{Typical Hamiltonian: weak coupling and low density}

We are now coming back to the initial physical motivations of Theorems
13, 16 and 17.  
These theorems show up quite strong conditions on the unitary operator 
$\LL=e^{-ihH}$ and a natural question now is: what kind of Hamiltonian
$H$ will produce such conditions on $\LL$? What is the typical
Hamiltonian for repeated quantum interactions which will produce
quantum Langevin equations at the continuous limit?

In this section we  answer partly that question. We answer it as
we exhibit a large family of such Hamiltonians and we conjecture that
they are the typical ones. We do not fully answer the question for we
are not able to prove that they are the only ones.
\bigskip
We keep here the notations of section II.1-II.3.

On $\rH_0\otimes\rH$ consider the following Hamiltonian
$$\di{
H=H_0\otimes I+I\otimes H_S+\frac{1}{\sqrt h}\sum_{i\in\L} \left(V_i\otimes
a^0_i+V_i^\ast \otimes a^i_0\right)
+\frac
1h\sum_{i,j\in\L} D_{ij}\otimes a^i_j \qq(15)\cr}
$$
where $H_0$, $V_i$ and $D_{ij}$ are bounded operators on $\rH_0$ (with
$H_0$ hermitian and $D_{ij}=D^*_{ji}$), and $H_S$ is bounded hermitian on $\rH$.

The condition that $H_S$ is bounded can be felt as a weakness in our
conditions. But one has to keep in mind that $\rH$ is just a ``small
piece'' of the bath system; it is only $\otimes_{\Rp}\rH$ which
represents the bath. In general $\rH$ is finite dimensional and the
resulting continuous field is a Fock space.
\smallskip
Put $\rH'$ to be the closed subspace of $\rH$ generated by the basis
elements $X^i$, $i\in\L$; that is, the orthogonal of
$X^0=\O$. Consider the ``column operator''
$$
V=\left(\matrix{V_1\cr V_2\cr\vdots\cr}\right)
$$
as an operator from $\rH_0$ to $\rH_0\otimes \rH'$.
Assume that this operator is bounded.

The adjoint of $V$ is then the ``row operator''
$$
V^\ast=\left(\matrix{V_1&V_2&\ld\cr}\right)
$$
from $\rH_0\otimes \rH'$ to $\rH_0$.

Define the ``matrix operator'' $D={(D^i_j)}_{ij}$ as an operator from
$\rH_0\otimes \rH'$ to $\rH_0\otimes \rH'$.
We also assume $D$ to be bounded.
\bigskip
Let us relate the above Hamiltonian with the usual literature on weak
coupling and low density limits.

Recall that, in the literature about weak coupling limit, the  bath is usually made of
several harmonic 
oscillators  (a Fock space) with associated creation operators
$a^\ast(g)$ and annihilation operators $a(g)$, where $g$ runs over a
Hilbert space $\rH$. The Hamiltonian which is considered is then of
the form
$$
H=H_0\otimes I+I\otimes H_S+\lambda\left(V\otimes
a^\ast(g)+V^\ast \otimes a(g)\right)
$$
The part $V\otimes
a^\ast(g)+V^\ast \otimes a(g)$ corresponds to the typical
dipole Hamiltonian usually considered in the weak coupling limit or van
Hove limit (cf [Dav], [D-J]).

This Hamiltonian meets (15) when considering an
orthonormal basis $(e_i)$ of $\rH$ and when one writes
$$
\left(V\otimes
a^\ast(g)+V^\ast \otimes a(g)\right)=\sum_i \left(\ps{e_i}g V\otimes a^\ast(e_i)+
\ps g{e_i} V^\ast\otimes a(e_i)\right)
$$ 
Our time renormalization term $1/\sqrt
h$ corresponds to the usual time renormalisation for the weak coupling limit
($t/\lambda^2=\tau$).
\bigskip
On the other hand, interaction Hamiltonians of the form
 $D\otimes a^\ast(f)a(h)+D^\ast\otimes
a^\ast(h)a(f)$ are typical of the low density limit (cf [APV],
[AFL]). They meet (15) when decomposing in the same way as above:
$$
\di{
\left(D\otimes a^\ast(f)a(h)+D^\ast\otimes
a^\ast(h)a(f)\right)=\sum_{i,j}\left(\ps{e_i}f\,\ps h{e_j}D\otimes
a^\ast(e_i)a(e_j)\right.+\hf\cr
\hf+\left.\ps f{e_i}\,\ps{e_j}hD^\ast\otimes
a^\ast(e_j)a(e_i)\right).\cr
}
$$
Our  time renormalization $1/h$ is also the typical one for
this limit.
\bigskip
We are now back to our general Hamiltonian (15).

We shall take compact notations for the unitary quantum Langevin
equations of Theorem 12. Consider the equation
$$
dU_t=\sum_{i,j}L^i_jU_t\, da^i_j(t)\eqno(16)
$$
with
$$
\eq{
L^0_0&=-(iK+\frac 12\sum_{k\in\L}L^\ast_kL_k)\cr
L^0_j&=L_j \vphantom{\sum_{k\in\L}}\cr
L^i_0&=-\sum_{k\in\L}L^\ast_kS^k_i\cr
L^i_j&=S^i_j-\d_{ij}I,\cr
}
$$
where $K$ is a bounded self-adjoint operator and $(S^i_j)_{i,j \in
\L}$ is unitary.

We write $W$ for the column operator
$$\left(\matrix{L_1\cr L_2\cr\vdots\cr}\right)$$and $S$ for the matrix
operator ${(S^i_j)}_{i,j}$. Then, with obvious notations
$$
\eq{
L^0_0&=-(iK+\frac 12 W^\ast W)\cr
L^0_\cd&=W\cr
L^\cd_0&=-W^\ast S\cr
L^\cd_\cd&= S-I.\cr
}
$$
With the Hamiltonian $H$ given by (15), put 
$$
U=e^{-ihH}
$$
and consider the evolution equation for repeated interactions
associated to $U$:
$$
u_{n+1}=U_{n+1}u_n.
$$

\th{19.}{\it The solution $\seqe u$ of the discrete time evolution
equation converges strongly in $\F$ to the solution $\pro U$ of the
quantum Langevin equation
$$
dU_t=\sum_{i,j}L^i_jU_t\, da^i_j(t)
$$
where, with the same notations as above
$$
\eq{
K&=H_0+\ps\O{H_S\O}I+V^\ast D^{-2}(\sin D-D)V\cr
W&=D^{-1}(e^{-iD}-I)V\cr
S&=e^{-iD}.\cr
}
$$
Moreover, for any bounded operator $X$ on $\rH_0 \otimes \Phi$, $u_n^*
X u_n$ converges to $U_t^* X U_t$.}
\bigskip
The convergences are meant, as in Corollary 18, for almost all $t$. The expressions $ D^{-2}(\sin D-D)$ and
$D^{-1}(e^{iD}-I)$ have to be understood has a short notation for the
associated convergent series, even if $D$ is not invertible.

\prf
Put 
$
k^i_j=\ps{X^j}{H_SX^i}
$
for all $i,j\in\L\cup\{0\}. $
We consider the column operator
$$
\vert\, k\rangle=\left(\matrix{k^0_1I\cr k^0_2I\cr\vdots\cr}\right),
$$
the row operator
$$
\langle k\, \vert=\left(\matrix{k^1_0I& k^2_0I&\ldots\cr}\right)
$$
and the matrix operator
$$
k={(k^i_jI)}_{i,j\in\L}\,.
$$
They all are bounded operators.
\bigskip
Put $\wt H=H_0+k^0_0I$ and $M={(M^i_j)}_{i,j\in\L}$ with
$M^i_j=\d_{ij}H_0+k^i_jI$. We then have 
$$
H=\left(\matrix{\wt H&\frac1{\sqrt h}V^\ast+\langle k\,\vert\cr
\frac 1{\sqrt h}\ecarte V+\vert\, k\rangle&\frac 1h D+M\cr}\right)
$$
as an operator on $\rH_0\otimes\rH$ which is  decomposed as an
operator on $\rH_0\otimes(\CC\O\oplus\rH')$. In particular
$$
hH=\left(\matrix{h\wt H&{\sqrt h}V^\ast+h\langle k\,\vert\cr
{\sqrt h}V+h\vert\,\ecarte k\rangle& D+hM\cr}\right)\, .
$$
Let $\a$ be a bound for $\norme {\wt H}$, $\norme V$, $\norme D$,
$\norme M$, $\norme{\langle k\, \vert}$, $\norme{\vert \, k\rangle}$.

\le{20.}{\it For all $m\in \NN$ we have
$$
{(hH)}^m=\left(\matrix{hA_m+h^{3/2}R^1_m&\sqrt hB_m+hR^2_m\cr
\sqrt hC_m+hR^3_m&\ecarte D_m+hR^4_m\cr}\right)
$$
with
$$\norme{X_m}\leq \a^m
$$
for all $X=A,B,C,D$ and
$$
\norme{R^i_m}\leq 7^{m-1}\a^m
$$
for all $i=1,2,3,4$; with
$$
\eq{
A_{m+1}&=V^\ast C_m,\qq A_0=I, A_1=\wt H\cr
B_{m+1}&=V^\ast D_m,\qq B_0=0, B_1=V^\ast\cr
C_{m+1}&=DC_m,\qq C_0=0, C_1=V\cr
D_{m+1}&=DD_m,\qq D_0=I.\cr}
$$
}

\noindent{\bf Proof of the lemma}

For $m=0$ the statements are clearly satisfied. For $m=1$ we find the
announced $A_1,B_1,C_1,D_1$ and $R^1_1=0$, $R^2_1=\langle k\, \vert$,
$R^3_1=\vert\, k\rangle$, $R^4_1=M$. The norm estimates are then clearly
satisfied.
 Now, applying the induction hypothesis and computing $(hH)^{m+1}$
from $(hH)^m$ we get 
$$
\eq{
A_{m+1}&=V^\ast C_m\qq \hbox{and\ }\norme{A_{m+1}}\leq \a\a^m=\a^{m+1}\cr
R^1_{m+1}&=\sqrt h\wt HA_m+h\wt HR^1_m+V^\ast
R^3_m+\langle k\, \vert C_m+\sqrt h\langle k\, \vert R^3_m\cr
\norme{R^1_{m+1}}&\leq 5(7^{m-1}\a^{m+1})\leq 7^{m}\a^{m+1}\cr
B_{m+1}&=V^\ast D_m\qq \hbox{and\ }\norme{B_{m+1}}\leq
\a\a^m=\a^{m+1}\cr
R^2_{m+1}&=\sqrt h\wt HB_m+h\wt HR^2_m+\sqrt hV^\ast R^4_m+\langle k\,
\vert D_m+h\langle k\, 
\vert R^4_m\cr
\norme{R^2_{m+1}}&\leq 5(7^{m-1}\a^{m+1})\leq 7^{m}\a^{m+1}\cr
C_{m+1}&=DC_m\qq \hbox{and\ }\norme{C_{m+1}}\leq
\a\a^m=\a^{m+1}\cr
R^3_{m+1}&=\sqrt hVA_m+hVR^1_m+D R^3_m+\sqrt hMC_m+hM R^3_m\cr
\norme{R^3_{m+1}}&\leq 5(7^{m-1}\a^{m+1})\leq 7^{m}\a^{m+1}\cr
D_{m+1}&=DD_m\qq \hbox{and\ }\norme{D_{m+1}}\leq
\a\a^m=\a^{m+1}\cr
R^4_{m+1}&=VB_m+hVR^2_m+\sqrt h\vert\, k\rangle B_m+h\vert\, k\rangle R^2_m+D
R^4_m+MD_m+hM R^4_m\cr 
\norme{R^4_{m+1}}&\leq 7(7^{m-1}\a^{m+1})\leq 7^{m}\a^{m+1}.\cr
}
$$
The lemma is proved.

In particular, we get
$$
\eq{
D_m&=D^m\cr
B_m&=V^\ast D^{m-1}\cr
C_m&=D^{m-1}V\cr
A_m&=V^\ast D^{m-2}V.\cr
}
$$
By the norm estimates of Lemma 20 we have that the series $$\sum_m
\frac{(-i)^m}{m!}R^j_m$$ are norm convergent. Let us denote by $R^k$
their repective limit.

We get
$$
\di{
\LL=\sum_m \frac{{(-i)}^m}{m!}{(hH)}^m=\hf\cr
=\left(\matrix{I-ih\wt H+h\sum_{m=2}^\infty\frac{{(-i)}^m}{m!} V^\ast
D^{m-2}V&&
\sqrt h\sum_{m=1}^\infty\frac{{(-i)}^m}{m!}V^\ast D^{m-1}\cr
+\ecarte h^{3/2}R^1&&+hR^2\cr\cr
\sqrt h\sum_{m=1}^\infty\frac{{(-i)}^m}{m!}D^{m-1}V&&
I+\sum_{m=1}^\infty \frac{{(-i)}^m}{m!}D^m\cr
+h\ecarte R^3&&+hR^4_m\cr
}\right)\,.
\cr
}
$$
This gives 
$$
U=\left(\matrix{I-ih\wt H+h V^\ast
D^{-2}(e^{-iD}-I+iD)V&&
\sqrt hV^\ast D^{-1}(e^{-iD}-I)\cr
+\ecarte h^{3/2}R^1&&+hR^2\cr\cr
\sqrt hD^{-1}(e^{-iD}-I)V&&
I+(e^{-iD}-I)\cr
+h\ecarte R^3&&+hR^4_m\cr
}\right)\,.
$$
We are exactly in the conditions of Corollary 18 and we thus get the
strong convergence to the solution of 
$$
dU_t=\sum_{i,j}L^i_jU_t\, da^i_j(t)
$$
where
$$
\eq{
L^0_0&=-i\wt H+V^\ast
D^{-2}(e^{-iD}-I+iD)V\cr
L^0_\cd&=D^{-1}(e^{-iD}-I)V\cr
L^\cd_0&=V^\ast D^{-1}(e^{-iD}-I)\cr
L^\cd_\cd&=e^{-iD}-I.\cr
}
$$
If we put $W=L^0_\cd$ and $S=e^{-iD}$, we then get
$$
-W^\ast S=-V^\ast (e^{iD}-I)D^{-1}S=V^\ast D^{-1}(e^{-iD}-I)=L^\cd_0
$$
and 
$$
\eq{
-\frac 12 W^\ast W&=-\frac 12 V^\ast
 (e^{iD}-I)D^{-1}D^{-1}(e^{-iD}-I)V\cr
&=-\frac 12 V^\ast D^{-2}
 (\cos D-I)V.\cr
}
$$
This shows that
$$
L^0_0=-i\wt H-iV^\ast D^{-2}(\sin D-D)V-\frac 12W^\ast W.
$$
The theorem is proved.\qed
\bigskip
Let us interpret the above theorem in terms of weak coupling and low
density limit again.
If in the above Hamiltonian we  consider no term $D_{ij}$, that
is,
$$
H=H_0\otimes I+I\otimes H_S+\frac{1}{\sqrt h}\sum_{i\in\L} \left(V_i\otimes
a^0_i+V_i^\ast \otimes a^i_0\right)
$$
then we are in the usual situation of a weak coupling limit, with its
typical dipole interaction Hamiltonian. The quantum Langevin equation
we obtain in Theorem 16 then simplifies to
$$
\di{
dU_t=-\left(iH_0+i\ps\O{H_S\O}I+\frac12\sum_i V_i^\ast V_i\right)U_t\,
dt+\hf\cr
\hf+\sum_i
V_iU_t\, da^0_i(t)-\sum_i V_i^\ast U_t\, da^i_0(t).\cr
}
$$
This is also typical of the ``diffusion'' type of quantum Langevin
equation one can meet in the literature for this kind of limit. The
fact that only creation and annihilation quantum noises are involved
is here the quantum analogue of a classical stochastic differential
equation with Brownian noise. Note that if the $V_i$'s are
such that $V_i^\ast=-V_i$ then the above quantum Langevin equation becomes a
classical stochastic differential equation with Brownian noises:
$$
dU_t=-\left(iH_0+i\ps\O{H_S\O}I+\frac12\sum_i V_i^2\right)U_t\, dt+\sum_i
V_iU_t\, dW_i(t).
$$
We refer to [At1] for a complete discussion on the classical
stochastic interpretations of the quantum noises.
\bigskip
On the other hand, if in the  Hamiltonian we  consider no term $V_i$, that
is,
$$
H=H_0\otimes I+I\otimes H_S+
\frac 1h\sum_{i,j\in\L} D_{ij}\otimes a^i_j
$$
then we are in the usual situation of a low density limit, with its
typical scattering-type interaction Hamiltonian. The quantum Langevin equation
we obtain in Theorem 18 then simplifies to
$$
dU_t=-i(H_0+\ps\O{H_S\O}I)U_t\, dt+\sum_{i,j \in \L} (S^i_j-\d_{ij})U_t\, da^i_j(t)
$$
where $S=e^{-iD}$.

This is also typical of the ``Poisson'' type of quantum Langevin
equation one can meet in the literature for this kind of limit. The
fact that only ``exchange'' quantum noises are involved
is here the quantum analogue of a classical stochastic differential
equation with Poisson noises.
\bigskip
Our formalism enables us to handle these two types of limits in a
single setup, and this could not have been done at the classical stochastic calculus level. 
This seems to be the first time in the literature, for,
to our knowledge, weak coupling limits and low density limits have
always been considered separately, as very different objects.

The surprise in our setup in the apparition of the term
$$
V^\ast D^{-2}(\sin D-D)V
$$
only when both the limits are in presence in the Hamiltonian.

Indeed, the term 
$$
L^0_0=-(iK+\frac 12 W^\ast W)
$$
is the driving term of the dynamic associated to the quantum Langevin
equation. In some sense it is the generator of the dynamic on $\rH_0$. The part
$\frac 12 W^\ast W$ is representative of the dissipation from $\rH_0$
to the bath. The part $iK$ is an effective Hamiltonian on $\rH_0$. The
apparition of this new contribution $
V^\ast D^{-2}(\sin D-D)V$ is new, and we have no
physical interpretation of it.

It is just clear that it results from the combined effects of the
weak coupling limit and the low density limit.

\sspa{IV.3}{Hamiltonian description of von Neumann measurements}

Our setup and approach allows to construct an Hamiltonian description
for the usual von Neumann measurement procedure (collapse of the wave
packet postulate). 

On some quantum system state space $\rH_0$ consider an observable $A$
with discrete spectrum (maybe infinite). Let $P_1,P_2,...$ denote the
associated spectral (orthogonal) projections, with $\sum_k P_k=I$.

We want to give a model for the action on $\rH_0$ of an exterior
measurement apparatus which measures the observable $A$. That is, the
action of the measurement apparatus on $\rH_0$ should be to transform
any observable $X$ of $\rH_0$ into 
$$
\sum_k P_k XP_k.
$$

Let $\rH$ be a Hilbert space with one more dimension than the number
of projectors $P_k$ involved above (infinite dimensional if the
$P_k$'s are in infinite number). On $\rH$ consider an orthonormal
basis $\O=X_0,X_1,X_2,...$ and the associated creation operators
$a^0_k$ and annihilation operators $a^k_0$, as in sections II.1-II.3.
Consider the following Hamiltonian on $\rH_0\otimes \rH$:
$$
H=\frac 1{\sqrt h}\sum_k \left(iP_k\otimes a^0_k-i P_k\otimes
a^k_0\right).
$$
Let $U=e^{-ihH}$ and consider the process 
$$
u_{n+1}=\LL_{n+1}u_n
$$
of repeated quantum
interactions on $\rH_0\otimes\bigotimes_{\NNE}\rH$.

\th{21.}{\it The repeated quantum interaction process $\seqe u$
converges strongly in $\rH_0\otimes\bigotimes_{\Rp}\rH$, when $h$ tends to 0,  to
the solution $\pro U$ 
of the quantum Langevin equation 
$$
dU_t=-\frac 12 U_t\, dt+\sum_k \left(iP_kU_t\, da^0_k(t)-iP_kU_t\,
da^k_0(t)\right).
$$
Furthermore, for any observable $X$ on $\rH_0$, the partial trace
along the field $\bigotimes_{\Rp}\rH$, in the vacuum state, of $U_t^\ast
(X\otimes I) U_t$ converges, when $t$ tends to $+\infty$, to 
$$
\sum_k P_kXP_k.
$$
}
\prf
In the basis $\O=X_0,X_1,X_2,...$ for $\rH$ the Hamiltonian $H$ writes
as
$$
H=\left(\matrix{0&-i\frac 1{\sqrt h}P_1&-i\frac 1{\sqrt h}P_2&\ld\cr
i\frac 1{\sqrt h}P_1&\ecarte 0&0&\ld\cr
i\frac 1{\sqrt h}P_2&\ecarte 0&0&\ld\cr
\vdots&\vdots&\vdots&\ddots\cr}\right).
$$
In particular, as an easy direct computation shows, we have
$$
\LL=\left(\matrix{\cos\sqrt h\, I&-\sin\sqrt h\, P_1&-\sin\sqrt h\, P_2&\ld\cr
\sin\sqrt h\, P_1&\ecarte\cos\sqrt h\, P_1&0&\ld\cr
\sin\sqrt h\, P_2&0&\ecarte\cos\sqrt h\, P_2&\ld\cr
\vdots&\vdots&\vdots&\ddots\cr}
\right).
$$
This unitary matrix clearly satisfies the conditions of Corollary 18 (one
could also have directly applied Theorem 19 to the Hamiltonian), we
thus have the strong convergence to the solution of the announced quantum Langevin
equation
$$
dU_t=-\frac 12 U_t\, dt+\sum_k \left(iP_kU_t\, da^0_k(t)-iP_kU_t\,
da^k_0(t)\right),
$$
with $U_0=I$.

Now, consider the evolution under $\pro U$ of a system observable $X$:
$$
U_t^\ast(X\otimes I)U_t
$$
and the partial trace along the field $\bigotimes_{\Rp}\rH$ in the vacuum
state:
$$
P_t(X)=\ps\O{U_t^\ast (X\otimes I)U_t\O}
$$
with the notation
$$
\ps{a}{\ps{\O}{A\O}\,b}_{\rH_0}=\ps{a\otimes\O}{A
\,(b\otimes\O)}_{\rH_0\otimes\bigotimes_{\Rp}\rH}
$$
for any operator $A$ on $\rH_0\otimes\bigotimes_{\Rp}\rH$, any $a,b\in\rH_0$.

It is well known form the usual theory of quantum stochastic
differential equations (cf [H-P]), that $\pro P$ is then a semigroup,
on $\rB(\rH_0)$, 
of completely positive maps with Lindblad generator 
$$
\rL(X)=-\frac 12\sum_k \left(P_kX+XP_k-2P_kXP_k\right).
$$
One could also have obtained this ``quantum master equation'' by simply
computing the discrete time quantum master equation associated to
$\seqe u$, using Theorem 2 and then by passing to the limit $h\rightarrow
0$ to recover $\pro P$ (see section IV.5). Our approach has the advantage to also describe the
exact equation for the interaction with the bath.

We thus get
$$
\rL(X)=\sum_k P_kXP_k-X
$$
so that
$$
\rL^2(X)=-\rL(X)
$$
and
$$
P_t(X)=e^{t\rL}(X)=\left(I+(1-e^{-t})\rL\right)(X).
$$
That is,
$$
P_t(X)=(1-e^{-t})\sum_kP_kXP_k+e^{-t}X
$$
which converges to 
$$
\sum_kP_kXP_k
$$
when $t$ tends to $+\infty$. \qed
\bigskip
We thus have proved the a von Neumann measurement apparatus can be
described in an Hamiltonian setup by, first considering an Hamiltonian
description of a repeated quantum interaction, secondly passing to the
limit to continuous quantum interactions ($h\rightarrow 0$) and thirdly passing to the limit to
large times ($t\rightarrow+\infty$).

\sspa{IV.4}{One example}
We  shall here follow a very basic example. It is
actually the simplest non-trivial physical example and it already
gives very interesting consequences.
\bigskip
 Assume $\rH_0=\rH=\CC^2$ that is, both are two-level systems with
basis states $\O$ (the fundamental state) and $X$ (the excited
state).

During the small amount of time $h$ the two systems are in contact and
they evolve in the following way:
\smallskip
-- if the states of the two systems are the same (both fundamental or
both excited) then nothing happens;
\smallskip
-- if they are different (one fundamental and the other one excited) then
they can  either be exchanged  or stay as they are.
\bigskip
In the basis $\{\O\otimes\O,\O\otimes X, X\otimes \O, X\otimes X\}$ the
operator $\LL$ is taken to be of the form
$$
\LL=\left(\matrix{1&0&0&0\cr
0&\cos\a&-\sin\a&0\cr0&\sin\a&\cos\a&0\cr0&0&0&1\cr}\right).
$$

The associated Hamiltonian is thus
$$
H=\left(\matrix{0&0&0&0\cr
0&0&-i\a/h&0\cr0&i\a/h&0&0\cr0&0&0&0\cr}\right)
$$
so that $\LL=e^{-ihH}$. 
\bigskip
For the choice $\a=\sqrt h$, that is,
$$
H=\frac{1}{\sqrt h}\left(\matrix{0&0&0&0\cr
0&0&-i&0\cr0&i&0&0\cr0&0&0&0\cr}\right)
$$
we get
$$
\LL=\left(\matrix{1&0&0&0\cr
0&\cos\sqrt h&-\sin\sqrt h&0\cr0&\sin\sqrt h&\cos\sqrt h&0\cr0&0&0&1\cr}\right).
$$
\bigskip
Repeating this interaction leads to an Hamiltonian equation of the
form
$$
u_{n+1}=\LL_{n+1}u_n
$$
with coefficients
$$
\LL^0_0=\left(\matrix{1&0\cr 0&\cos\sqrt h}\right),\ 
\LL^0_1=\left(\matrix{0&\sin\sqrt h\cr 0&0}\right),
$$
$$
\LL^1_0=\left(\matrix{0&0\cr -\sin\sqrt h&0}\right),\ 
\LL^1_1=\left(\matrix{\cos\sqrt h&0\cr 0&1}\right)
$$
for $\LL$.  
We then have
$$
\eq{
\lim_{h\rightarrow 0}\frac{
\LL^0_0-I}h&=\left(\matrix{0&0\cr0&-1/2}\right)\cr
\lim_{h\rightarrow 0}\frac{
\LL^0_1}{\sqrt h}&=\left(\matrix{0&1\cr0&0}\right)\cr
\lim_{h\rightarrow 0}\frac{
\LL^1_0}{\sqrt h}&=\left(\matrix{0&0\cr -1&0}\right)\cr
\lim_{h\rightarrow 0}{
\LL^1_1-I}&=\left(\matrix{0&0\cr0&0}\right).\cr
}
$$
We are exactly in the condition for applying Theorem 15 since for all
$h$ the matrix $\LL (h)$ is unitary. We get that for almost all $t$,
$u_{[t/h]}$ converges strongly to $U_t$ where $(U_t)_{t\in \RR_+}$ is the
unitary solution of 
$$ dU_t = -\frac12 V^*V \, U_t \, dt + V U_t \, da^0_1 (t) - V^\ast U_t \, da^1_0 (t)$$
with $V= \left( \matrix{0&1\cr0&0}\right)$. This equation  is the
well-known quantum Langevin equation
associated to the spontaneous decay into the ground state in the
Wigner-Weisskopf model for the two-level atom (see [M-R] for example).

\sspa{IV.5}{From completely positive maps to Lindbladians}

Recall, from section II.3 that the solution $\seq u$ of the equation
$$
u_{n+1}=\LL_{n+1}u_n
$$
with $u_0=I$ induces a completely positive evolution on the small
system. Namely, in the Heisenberg picture, one has for any $a,b$ in
$\rH_0$, any $X$ in $\rB (\rH_0)$,
$$
\ps{a\otimes\O}{u_n^\ast X u_n \, b\otimes\O}=\ps{a}{\ell^n(X) \, b}
$$
where 
$$
\ell(X)=\sum_{i\in\NN} \,( \LL^0_i)^\ast X  \LL^0_i.
$$

\th{22.}{\it Let $\LL(h)=\left(\LL^i_j(h)\right)_{i,j=0,\ld n} $ be a
family of matrices such that $\sum_i\LL^0_i(h)^\ast  \LL^0_i(h) =I$ and such
that
\smallskip
-- $(\LL^0_0(h)-I)/h $ converges to some $L^0_0$,
\smallskip
-- $\LL^0_i/\sqrt h$ converges to some $L^0_i$ for all $i=1, \ld, n$.
\smallskip
Then there exists a self-adjoint operator $H$ in $\rB(\rH_0)$ such
that for all $t\in\Rp$,
$$
\ell^{[t/h]}\longrightarrow e^{t\rL}
$$
in operator norm, where $\rL$ is the Lindblad generator
$$
\rL(X)=i[H,X]+\frac12\sum_{i\in\N}\left(2 {L^0_i}^\ast X {L^0_i} - 
{L^0_i}^\ast {L^0_i} X - X {L^0_i}^\ast {L^0_i}\right)
$$
and $\ell$ is defined above.}

\prf

It is a straightforward computation that 
$$
\ell(X)=X+h({ L^0_0}^\ast X  + X {L^0_0} )+h\sum_{i=1}^n 
{L^0_i}^\ast X {L^0_i} +o(h\norme X).
$$
The equality $\ell(I)=I$ entails
$$
({L^0_0}^\ast+L^0_0)+\sum_{i\in\NN} {L^0_i}^\ast {L^0_i} =0
$$
so that
$$
i\left(L^0_0+\frac 12\sum_{i\in\NN} { L^0_i}^\ast {L^0_i} \right)
$$ is self-adjoint. We denote it by $H$. Then $\ell$ is of the form
$$
\ell(X)=X+h\left(i[H,X]+\frac12\sum_{i\in\NN}\left(2 {L^0_i}^\ast X {L^0_i} - 
{L^0_i}^\ast {L^0_i} X- X {L^0_i}^\ast {L^0_i} \right)\right)+o(h\norme X),
$$
which is, with the notations of the statements
$$
\ell(X)=X+h\rL(X)+o(h\norme X).
$$
The above mentioned convergence is therefore clear.\qed
\bigskip
As a consequence, it is very easy to obtain approximations of
solutions of continuous-time master equations: 
$$
\frac{dX_t}{dt}=\rL(X_t)
$$
by solutions of discrete-time ones:
$$
x_{n+1}=\ell(x_n).
$$
Yet notice that the master equation gives no information whatsoever on
the interaction between the small system and the environment or on the
environment itself. On the other hand, the associated quantum Langevin
equation contains the information of the whole system; this justifies
our effort.

Notice that, in the above proposition, no hypothesis is needed on the
other coefficients of the matrix $\LL(h)$. Their properties actually
depend on the choice of additional features of the matrices $\LL(h)$,
for example their unitarity. The possibility of choosing $\LL(h)$ to
be unitary and obtain in the end the desired Lindbladian ${\cal L}$ is
described by Parthasarathy in exercises 29.12 and 29.13 of [Par].

What's more, these manipulations show that the hypotheses of
convergence of Theorem 13 are not as artificial as it seems, and are
not only convenient assumptions we set up in order to obtain the right
convergence. Indeed, to $\LL(h)$ is associated both a dynamic on the
observables, as we have seen, and an evolution $\tau$, defined by
 
$$ \ps {a}{\tau _n b} = \ps{a\otimes \O}{u_n \, b\otimes \O}$$
for all $a$,$b$ in $\rH_0$, which turns our to be
$$ \tau _n = (\LL^0_0) {}^n.$$
If one assumes that $\tau_{[t/h]}$ converges for almost all $t$ and
that $\LL^0_0$ is assumed to be continuous at $h=0$ then the
assumption on $\LL^0_0$ in Theorem 16 is to be fulfilled; this implies
that $\sum _{i\in\L}\LL^i_0{}^*   \LL^i_0  = -h (L^0_0 {}^* + L^0_0) +
o(h)$, so that the other assumptions of convergence of Theorem 13 are
natural.

The other conditions described in Theorem 17 are in turn necessary if
one wants the process $\pro U$ obtained in the limit to be unitary
or alternatively the matrices $\LL (h)$ to be sufficiently close to unitarity.

\vfill\eject
\spa{}{References}
  
\bigskip\noindent\hangindent=1cm\hangafter=1
[AFL]: Accardi L., Frigerio A., Lu Y.G. 
 ``Quantum Langevin equation in the weak coupling limit'', 
 {\it Quantum probability and applications} V (Heidelberg, 1988), 1--16, 
 Lecture Notes in Math., 1442, 
 Springer, Berlin, 1990. 
 \bigskip\noindent\hangindent=1cm\hangafter=1
[AGL]: Accardi L., Gough J., Lu Y. G.
 ``On the stochastic limit for quantum theory'', 
 {\it Proceedings of the XXVII Symposium on Mathematical Physics} (Toru\'n, 1994). 
 {\it Rep. Math. Phys.} 36 (1995), no. 2-3, 155--187.
\bigskip\noindent\hangindent=1cm\hangafter=1
[ALV]: Accardi L., Lu Y. G., Volovich I. 
 ``{\it Quantum theory and its stochastic limit}'',
 Springer-Verlag, Berlin, 2002. 
\bigskip\noindent\hangindent=1cm\hangafter=1
[A-L]: Accardi L., Lu Y. G.
 ``The low density limit and the quantum Poisson process'', 
 {\it Probability theory and mathematical statistics}, Vol. I
 (Vilnius, 1989), 1--27,  
 ``Mokslas", Vilnius, 1990. 
\bigskip\noindent\hangindent=1cm\hangafter=1
[APV]: Accardi L., Pechen A.N., Volovich I.V. ``A stochastic golden
rule and quantum Langevin equation for low density limit'', {\it preprint}
\bigskip\noindent\hangindent=1cm\hangafter=1
[At1]: Attal S. 
 ``Extensions of the quantum stochastic calculus''
{\it  Quantum Probability Communications} XI, 1-38, 
 World Scientific,  2003. 
\bigskip\noindent\hangindent=1cm\hangafter=1
[At2]: Attal S. 
 ``Approximating the Fock space with the toy Fock space''
{\it  S\'eminaire de Probabilit\'es} XXXVI, 477--491, 
 Lecture Notes in Math., 1801, 
 Springer, Berlin, 2003.
 \bigskip\noindent\hangindent=1cm\hangafter=1
[At3]: Attal S. 
 ``{\it Quantum noise theory, with applications to stochastic calculus
and quantum open systems}'', book in preparation.
\bigskip\noindent\hangindent=1cm\hangafter=1
[A-L]:  Attal S., Lindsay J.M. 
 ``Quantum stochastic calculus with maximal operator domains'',
 {\it The Annals of Probability}, to appear.
\bigskip\noindent\hangindent=1cm\hangafter=1
[A-M]:  Attal S., Meyer P.-A. 
 ``Interpr\'etation probabiliste et extension des int\'e\-gra\-les stochastiques
 non commutatives'',
 {\it S\'eminaire de Probabilit\'es} XXVII, 312--327, 
 Lecture Notes in Math., 1557, 
 Springer, Berlin, 1993. 
\bigskip\noindent\hangindent=1cm\hangafter=1
[A-P]:  Attal S., Pautrat Y. 
 ``From $(n+1)$-level atom chains to $n$ dimensional noises'',
 {\it preprint}
\bigskip\noindent\hangindent=1cm\hangafter=1 
[Ba1]: Barchielli A. 
 ``Measurement theory and stochastic differential equations in quantum
 mechanics'',  
 {\it Phys. Rev. A} (3) 34 (1986), no. 3, 1642--1649.
 \bigskip\noindent\hangindent=1cm\hangafter=1
 [Ba2]: Barchielli A. 
 ``Some stochastic differential equations in quantum optics and
 measurement theory: the case of counting processes'',
 {\it Stochastic evolution of quantum states in open systems and in
 measurement processes} (Budapest, 1993), 1--14,  
 World Sci. Publishing, River Edge, NJ, 1994. 
 \bigskip\noindent\hangindent=1cm\hangafter=1
[B-B]: Barchielli A., Belavkin V.P.
 ``Measurements continuous in time and a posteriori states in quantum mechanics'', 
 {\it J. Phys. A} 24 (1991), no. 7, 1495--1514.
\bigskip\noindent\hangindent=1cm\hangafter=1
[BRSW]: Bellissard J., Rebolledo R., Spehner D., von Waldenfels W. 
``The Quantum Flow of Electronic transport'',  {\it preprint}. 
\bigskip\noindent\hangindent=1cm\hangafter=1
[Coq]: Coquio A.,
 ``Why are there only three quantum noises?'', 
 {\it Probab. Theory Related Fields} 118 (2000), no. 3, 349--364.
\bigskip\noindent\hangindent=1cm\hangafter=1
[Dav]: Davies E.B., 
 Markovian master equations. 
 {\it Comm. Math. Phys.} 39 (1974), 91--110.  
\bigskip\noindent\hangindent=1cm\hangafter=1
[D-J]: Derezinski J., Jaksic V. ``On the nature of Fermi Golden Rule
for open quantum systems'',  {\it J. Stat. Phys.}, to appear.
\bigskip\noindent\hangindent=1cm\hangafter=1
[F-R]: Fagnola F., Rebolledo R.
 ``A view on stochastic differential equations derived from quantum
 optics'',  
 Stochastic models  (Guanajuato, 1998), 193--214, 
 {\it Aportaciones Mat. Investig., 14, 
 Soc. Mat. Mexicana}, México, 1998. 
 \bigskip\noindent\hangindent=1cm\hangafter=1
[FRS]: Fagnola F., Rebolledo R., Saavedra C. 
 ``Quantum flows associated to master equations in quantum
 optics'',  
 {\it J. Math. Phys.} 35 (1994), no. 1, 1--12.
 \bigskip\noindent\hangindent=1cm\hangafter=1
[FKM]: Ford G. W., Kac M., Mazur P. 
 ``Statistical mechanics of assemblies of coupled oscillators'', 
 {\it J. Mathematical Phys.} 6 1965 504--515.
 \bigskip\noindent\hangindent=1cm\hangafter=1
[FLO]:  Ford G. W., Lewis J. T., O'Connell R. F.
 ``Quantum Langevin equation'', 
 {\it Phys. Rev. A} (3) 37 (1988), no. 11, 4419--4428.
\bigskip\noindent\hangindent=1cm\hangafter=1
[G-Z]: Gardiner C. W., Zoller P.       
 {\it Quantum noise.
 A handbook of Markovian and non-Markovian quantum stochastic methods
 with applications to quantum optics.}, Second edition. Springer Series
 in Synergetics.  
 Springer-Verlag, Berlin, 2000. 
\bigskip\noindent\hangindent=1cm\hangafter=1 
[GSI]: Grabert H., Schramm P., Ingold
 G-L  
 ``Quantum Brownian motion: the functional integral approach'', 
 {\it Phys. Rep.} 168 (1988), no. 3, 115--207.
\bigskip\noindent\hangindent=1cm\hangafter=1 
[Gui]: Guichardet A.
 {\it Symmetric Hilbert spaces and related topics.},  Lecture Notes in Mathematics, Vol. 261. 
 Springer-Verlag, Berlin-New York, 1972.
 \bigskip\noindent\hangindent=1cm\hangafter=1
[H-P]: Hudson R. L., Parthasarathy K. R.
 ``Quantum Ito's formula and stochastic evolutions'',
 {\it Comm. Math. Phys.} 93 (1984), no. 3, 301--323.
\bigskip\noindent\hangindent=1cm\hangafter=1
[L-M]: Lindsay J.M., Maassen H. 
 ``Stochastic calculus for quantum Brownian motion of nonminimal
variance---an approach using integral-sum kernel operators'',  
{\it  Mark Kac Seminar on Probability and Physics} Syllabus 1987--1992
(Amsterdam, 1987--1992), 97--167,  
 CWI Syllabi, 32, 
 Math. Centrum, Centrum Wisk. Inform., Amsterdam, 1992. 
 \bigskip\noindent\hangindent=1cm\hangafter=1
[M-R]: Maassen H., Robinson P. 
 ``Quantum stochastic calculus and the dynamical Stark effect'', 
 {\it Rep. Math. Phys.} 30 (1991), no. 2, 185--203 (1992).
\bigskip\noindent\hangindent=1cm\hangafter=1
[Par]: Parthasarathy K.R. 
 ``{\it An introduction to quantum stochastic calculus}'', Mo\-no\-graphs
in Mathematics 85, Birkh\"auser (1992).
 \bigskip\noindent\hangindent=1cm\hangafter=1
[Pa1]: Pautrat Y., ``From Pauli matrices to quantum Ito formula'', 
{\it preprint}.
 \bigskip\noindent\hangindent=1cm\hangafter=1
[Pa2]: Pautrat Y., ``Kernel and integral representations of operators
on  infinite dimensional toy Fock space'', 
 {\it S\'eminaire de Probabilit\'es}, to appear. 
 \bigskip\noindent\hangindent=1cm\hangafter=1
[Pa3]: Pautrat Y., ``Des matrices de Pauli aux bruits quantiques",
{\it Th\`ese de doctorat de l'Universit\'e Joseph Fourier} (2003).

\bigskip\bigskip
\line{St\'ephane Attal\hf}
\line{ Institut FOURIER, U.M.R. 5582\hf}

\line{ Universit\'e de Grenoble I, BP 74\hf }

\line{ 38402 St Martin d'H\`eres cedex, France\hf }

\bigskip
\line{Yan Pautrat\hf}
\line{ McGill mathematics and statistics\hf}

\line{ 805 Sherbrooke West\hf }

\line{ Montreal, QC, H3A 2K6, Canada\hf }

\end